%% file: main.tex
\documentclass[manuscript,screen]{acmart}

\usepackage{multirow}
\usepackage{cleveref}
\usepackage{colortbl}
\usepackage{tikz}
\usepackage{amsmath}
\usepackage[inkscapeformat=png]{svg}

\usepackage{filecontents}

\let\origwidebar\widebar 
\let\origbigtimes\bigtimes 
\let\bigtimes\relax 
\let\widebar\relax 
\usepackage{mathabx}
\let\bigtimes\origbigtimes 
\let\widebar\origwidebar 

\usepackage{wasysym}

\usepackage{subfig}
\usepackage{float}

\usepackage{csquotes}
\AtBeginDocument{%
  \providecommand\BibTeX{{%
    \normalfont B\kern-0.5em{\scshape i\kern-0.25em b}\kern-0.8em\TeX}}}

\setcopyright{acmcopyright}
\copyrightyear{2024}
\acmYear{2024}
\acmDOI{XXXXXXX.XXXXXXX}

\begin{document}

\title{On the Usability of Next-Generation Authentication: A Study on Eye Movement and Brainwave-based Mechanisms}

\renewcommand{\shorttitle}{On the Usability of Next-Generation Authentication}


\author{Matin Fallahi}
\affiliation{%
  \institution{KASTEL Security Research Labs, KIT}
  \city{Karlsruhe}
  \country{Germany}}
\email{matin.fallahi@kit.edu}

\author{Patricia Arias-Cabarcos}
\affiliation{%
  \institution{Paderborn University and KASTEL Security Research Labs}
  \city{Paderborn}
  \country{Germany}}
\email{pac@mail.upb.de}

\author{Thorsten Strufe}
\affiliation{%
 \institution{KASTEL Security Research Labs, KIT}
 \city{Karlsruhe}
 \state{Arunachal Pradesh}
 \country{Germany}}
\email{strufe@kit.edu}

\renewcommand{\shortauthors}{Fallahi and others}

\begin{abstract}
Passwords remain a widely-used authentication mechanism, despite their well-known security and usability limitations. To improve on this situation, next-generation authentication mechanisms, based on behavioral biometric factors such as eye movement and brainwave have emerged. However, their usability remains relatively under-explored. To fill this gap, we conducted an empirical user study (n=32 participants) to evaluate three brain-based and three eye-based authentication mechanisms, using both qualitative and quantitative methods. Our findings show good overall usability according to the System Usability Scale for both categories of mechanisms, with average SUS scores in the range of 78.6-79.6 and the best mechanisms rated with an ``excellent'' score.  Participants particularly identified brainwave authentication as more secure yet more privacy-invasive and effort-intensive compared to eye movement authentication. However, the significant number of neutral responses indicates participants' need for more detailed information about the security and privacy implications of these authentication methods. Building on the collected evidence, we identify three key areas for improvement: privacy, authentication interface design, and verification time. We offer recommendations for designers and developers to improve the usability and security of next-generation authentication mechanisms. 
\end{abstract}

\ccsdesc[500]{Security and privacy~Biometrics}

\keywords{User study, Usability, Authentication, EEG, Biometric, eye movement}



\maketitle

\input{01_introduction}

\input{02_relatedwork}
\input{03_Prototypes}

\input{04_methodology}

\input{09_result}

\input{10_discussion}

\section{Conclusion}
\label{sec:Conclusion}
In conclusion, our investigation into next-generation authentication methods based on behavioral biometrics, like eye movement and brainwave, reveals promising potential for user acceptance. Our study results on usability showed good to excellent scores in the System Usability Scale (SUS), with averages of 78.6 for eye mechanisms and  79.6 for brainwave-based. To improve usability, authentication tasks that require a low cognitive effort are preferred. When compared to the eye movement mechanisms, subjects perceive the brainwave mechanism to be more secure; however, they also express increased concerns regarding privacy and reliability of brainwaves, in both qualitative and quantitative questions. While attitudes are generally positive, concerns on privacy, security understanding, and efficient performance need to be further investigated.

\input{11_Ack}


\end{document}

%% file: 01_introduction.tex
\section{Introduction}
Authentication is a cornerstone of security, ensuring that only authorized individuals gain access to sensitive systems or data. However, traditional methods relying on single knowledge factors such as passwords and PINs have shown significant drawbacks. Recent studies, including the 2020 Data Breach Investigations Report by Verizon \cite{verizon2020}, emphasize that approximately 80\% of hacking-related breaches involve weak or stolen credentials, with passwords being a prime target. Furthermore, the Ponemon Institute revealed that 51\% of respondents admitted to using the same password for multiple accounts, consequently increasing the risks of credential theft and identity fraud \cite{ponemon2020}. These statistics shed light on the vulnerabilities inherent in password-based authentication systems, calling for more robust, usable, and secure alternatives.

One promising solution to address the limitations of traditional authentication methods is biometric authentication \cite{sarkar2020review}. This approach harnesses the unique physiological or behavioral characteristics of individuals to verify their identities. While physiological biometrics like face and fingerprint recognition are popular, they face critical challenges, including the inability to revoke biometric data once compromised and heightened vulnerability to spoofing attacks \cite{kumar2017comparative,arora2016design}. On the other hand, behavioral biometrics measure unique features of activities users perform either consciously or unconsciously \cite {ballard2007forgery}. Behavioral biometrics have gained significant attention as they offer the potential to enhance security while minimizing user burden \cite{stragapede2022mobile}.

Among the behavioral biometric authentication approaches, brainwave-based \cite{sooriyaarachchi2020musicid, schomp2018behavioral} and eye movement-based \cite{lohr2022eye} mechanisms\footnote{Brainwaves and eye movement patterns are generally categorized as behavioral biometrics~\cite{hogben2010enisa}, though they are also influenced by physiological aspects like the thickness of the skull or the dimensions of the eyeball. We conform to their categorization as behaviorals, which is dominant in the literature.} have emerged as promising alternatives in desktop and Extended Reality (XR) environments, due to their distinct advantages. These mechanisms allow for implicit authentication, without requiring explicit user actions, such as typing a password or pressing a button, ensuring a seamless and effortless authentication experience. Furthermore, brainwaves are non-observable from the exterior and therefore difficult to compromise, and brain biometrics can be implemented in a adaptable fashion by altering stimuli even if the original brainwave sample is compromised ~\cite{lin2018brain,wheeler2010face,korany2019xmodal}. Eye-based mechanisms do not require a wearable and can work with common camera hardware integrated into laptops/smartphones~\cite{yang2021webcam, liu2015exploiting}. Lastly, both types of mechanisms have demonstrated promising authentication accuracy in previous research \cite{sluganovic2018analysis, fallahi2023brainnet}, which supports their potential for practical realization in the near future.

Despite the potential of brainwave and eye movement-based authentication mechanisms, their actual usability remains under-explored, and this is a crucial factor driving user acceptance and influencing security in practice. The limited number of prior studies investigating these mechanisms lacked a standardized approach for assessing perceived usability, such as the System Usability Scale (SUS)~\cite{brooke1996sus}, and failed to provide usage conditions for an ecologically valid evaluation \cite{arias2023performance, chuang2013think,brooks2013perceptions}. The main barrier in this regard is the absence of real authentication prototypes integrating brain and eye-based authentication mechanisms, which has hindered comprehensive evaluations of their practicality. To bridge this gap, we aim at answering the following research questions:
 
 \begin{itemize}

    \item \textbf{RQ1 [Usability]} How usable are brainwave-based and eye movement-based authentication mechanisms as perceived by users?
    
    \item \textbf{RQ2 [Perceptions \&Usage]} How do users perceive brainwave-based and eye movement-based authentication mechanisms in terms of security, reliability, and effort? How would they use these mechanisms?

\item \textbf{RQ3 [Benefits, Problems, \& Tradeoffs]} What are the advantages, disadvantages, and tradeoffs of brainwave-based and eye movement-based authentication mechanisms from the users' perspective?
   
\end{itemize}
 
 To answer the above questions, we conducted a lab study with 32 participants (\Cref{sec:Methodology}). We tested three brainwave-based and three eye movement-based authentication methods in a controlled experiment. To facilitate early usability evaluation, protect user privacy, and ensure ethical research practices, our approach involved using interactive mock-ups that did not collect any actual biometric data as suggested by similar research~\cite{zimmermann2020password,khan2015usability, trewin2012biometric,rose2023overcoming}. Our mock-ups were designed to realistically simulate the authentication mechanisms in a real-world use-case scenario: authenticating to a news website. We collected quantitative and qualitative data from participants after interacting with our prototypes, including System Usability Scale (SUS) scores and responses to open-ended questions on envisioned benefits, problems, and other acceptability-related dimensions, i.e., privacy, confidence, and security. Our results show strong usability for eye and brain-based authentication and high intention to use them by the study participants.

Overall, the mock-up study served as an efficient proxy\footnote{Only 8.7\% of the participants suspected it was a simulation, and they were excluded from the results} for assessing the usability of behavioral authentication interfaces before implementing full prototypes and led us to identify three key areas of improvement: privacy and transparency, authentication interface design, and verification time. Beyond recommendations to target these issues, we contribute the prototypes as a blueprint framework for developers and researchers to conduct future work\footnote{\url{https://github.com/kit-ps/mockup_paper/}}.

%% file: 02_relatedwork.tex
\section{Related Work}
\label{sec:RW}

Several studies have investigated the \textbf{technical} and \textbf{user-perception} aspects of behavioral biometrics.

On the \textbf{technical studies} side, numerous works have evaluated the accuracy, efficiency, and security of behavioral authentication systems. For instance, \textit{Xu et al.}\cite{xu2020gait},\textit{ Lin et al.}\cite{lin2022crossbehaauth}, \textit{Fallahi et al.}~\cite{fallahi2023brainnet}, and\textit{ Eberz et al.} \cite{eberz201928} have examined the performance of various behavioral biometric modalities in terms of Equal Error Rates (EER), including gait (EER=3.5\%), keystroke dynamics (EER=5.35\%), brainwaves (EER=0.14\%), and eye movement (EER=1.88\%). The promising results of these studies suggest that behavioral biometrics have the potential to be integrated into daily life. However, a crucial aspect that requires investigation is users' perceptions of these authentication mechanisms.

On the \textbf{user perception} stream of research, most works have focused on examining factors that affect user acceptance and adoption of physiological biometric authentication rather than behavioral biometrics. These factors include perceived security \cite{zimmermann2020password}, privacy concerns \cite{riley2009culture}, perceived ease of use \cite{morosan2012voluntary}, and social influence \cite{de2015feel}. Several studies have shown that while users generally have a positive view of biometric authentication, they may still have concerns regarding privacy and data protection \cite{zimmermann2020password, furnell2007public, jones2007towards}. However, a limited number of laboratory studies have been conducted pertaining to behavioral biometrics in general and brainwaves and eye movement authentication in particular. The few studies in this area were constrained by the unavailability of real-world prototypes and their evaluations rely on hypothetical scenarios or partial interface elements rather than on actual user interaction with a functional biometric system. 
  
In the case of brainwave authentication, \textit{ Chuang et al.}~\cite{chuang2013think}, conducted a usability study asking participants (N=15) to rate authentication
tasks according to how enjoyable, easy, or engaging they were. Building on this study, \textit{Arias-Cabarcos et al.}~\cite{arias2023performance} provided a more comprehensive evaluation (N=52), extending the questionnaire to cover both the usability of the EEG device and explore attitudes towards acceptance. However, in both studies, participants only evaluated authentication tasks, i.e., the activity performed by users while measured by a neuroheadset in order to get identified (e.g., resting, moving a hand, looking at images). But tasks are only a detached part of the whole authentication experience, so the insights regarding their usability provided limited value. However, recent research by Röse et al.~\cite{rose2023overcoming} explored the usability of brainwave authentication by simulating its use as an authentication mechanism for a password manager. In their preliminary study, they achieved an excellent System Usability Scale (SUS) score of 85.28, based on surveys completed by 9 participants and simulating the ideal (but unrealistic) case where legitimate users are never rejected. Notably, this investigation was focused solely on a single authentication task, consisting of showing a slideshow of images.

In the case of eye movement-based authentication, the study by \textit{Brooks et al.} \cite{brooks2013perceptions} provides valuable insights into user perceptions (N=22) based on their interaction with a simulated biometric system. However, the simulation does not take into account the expected performance parameters of eye-based systems and the study did not capture the standardized system usability questionnaire (SUS), which may limit the ability to compare their results with other studies or assess the usability of the authentication schemes in a broader context. Additionally, while the study \cite{brooks2013perceptions} focused on other factors that may influence user acceptance, it did not specifically address privacy concerns, which is an important consideration for biometric authentication.

We complement and extend existing research by comprehensively evaluating brainwaves and eye movement behavioral biometric systems, building high-fidelity interactive prototypes for the most promising interfaces in the literature, and considering their theoretical performance to simulate authentication. Furthermore, we address issues related to various aspects of user perception, conduct standardized usability tests, and investigate privacy concerns in order to gain a clearer understanding of the usability and potential acceptance of behavioral authentication methods.

%% file: 03_Prototypes.tex
\section{Authentication Prototypes}

\label{sec:Prototypes}
To enable realistic testing of authentication mechanisms, we implemented a news website that required registration and subsequent authentication to get access to extended content. This scenario is common and familiar to users, as it resembles the usual flow such as face detection. Here, users are required to provide biometric samples at the time of enrollment and then submit new samples during verification to either accept or deny authentication requests.

\textbf{Authentication System Flow and Elements. }In a behavioral biometric authentication system, users are granted access depending on their distinct traits, such as those we set up to study: brain activity and gaze. The full process involves collecting the data through specific hardware sensors, processing these data to extract relevant features, and comparing them to a previously stored sample or template from the user trying to authenticate, checking if it is a match or a mismatch. To acquire brainwave and gaze data for behavioral authentication, users should perform a specific task or be presented with certain stimuli, such as sounds or images. In our prototypes, we have developed interfaces for the authentication tasks and employed a simulated authentication decision algorithm. To enhance the realism of our experiment, even in the absence of actual user data collection, participants interact with the necessary hardware components as they would in a fully operational biometric system. Details on the prototype are given in the following.

\begin{figure*}[!t]
  \centering
  \subfloat[Technical setup]{\label{F:emotiv}\includegraphics[width=0.5\textwidth]{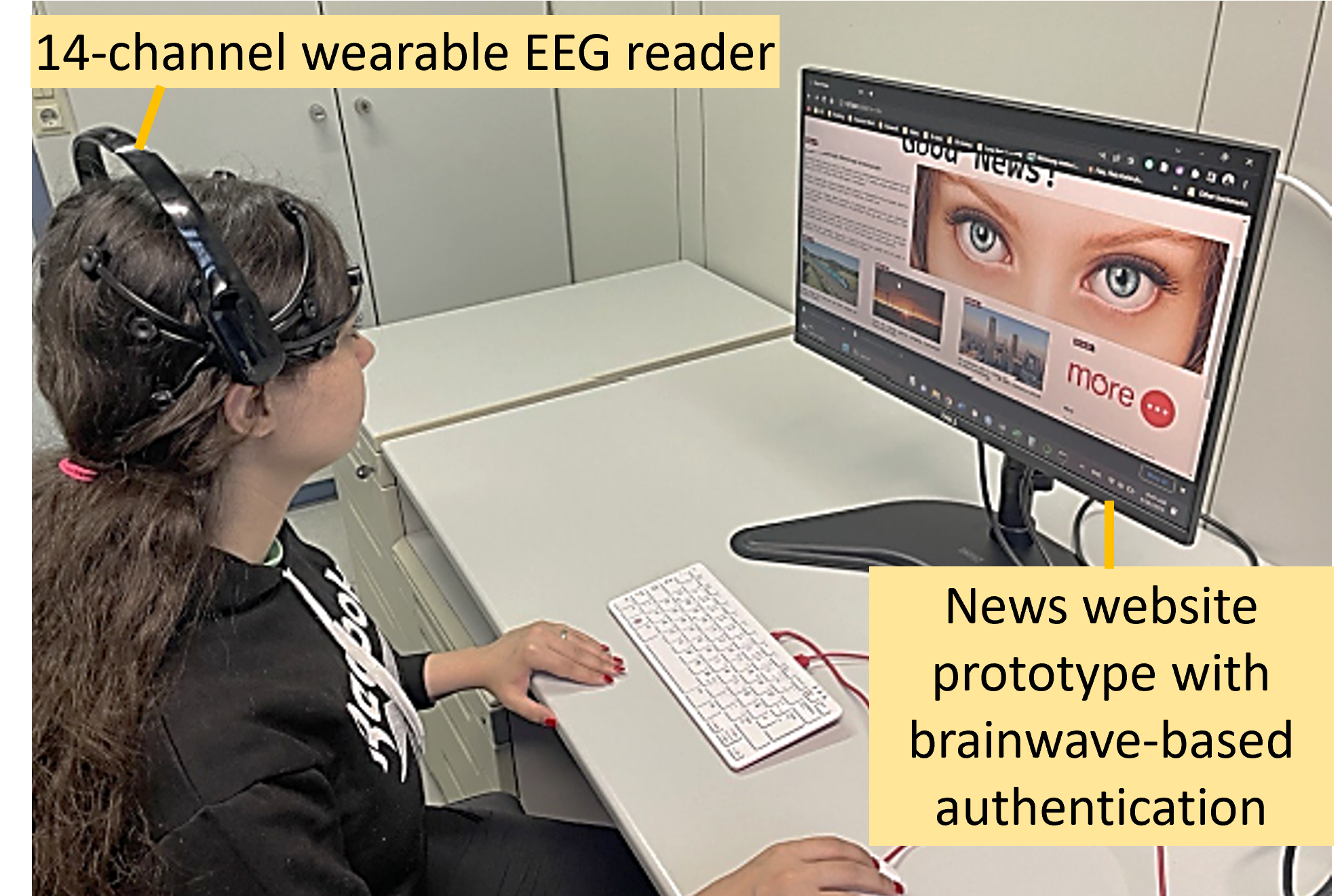}}  
\hspace{0.1cm}   
\subfloat[Authentication Interfaces]
  {\label{P:tasks}{\raisebox{0.85cm}{\includegraphics[width=0.45\textwidth]{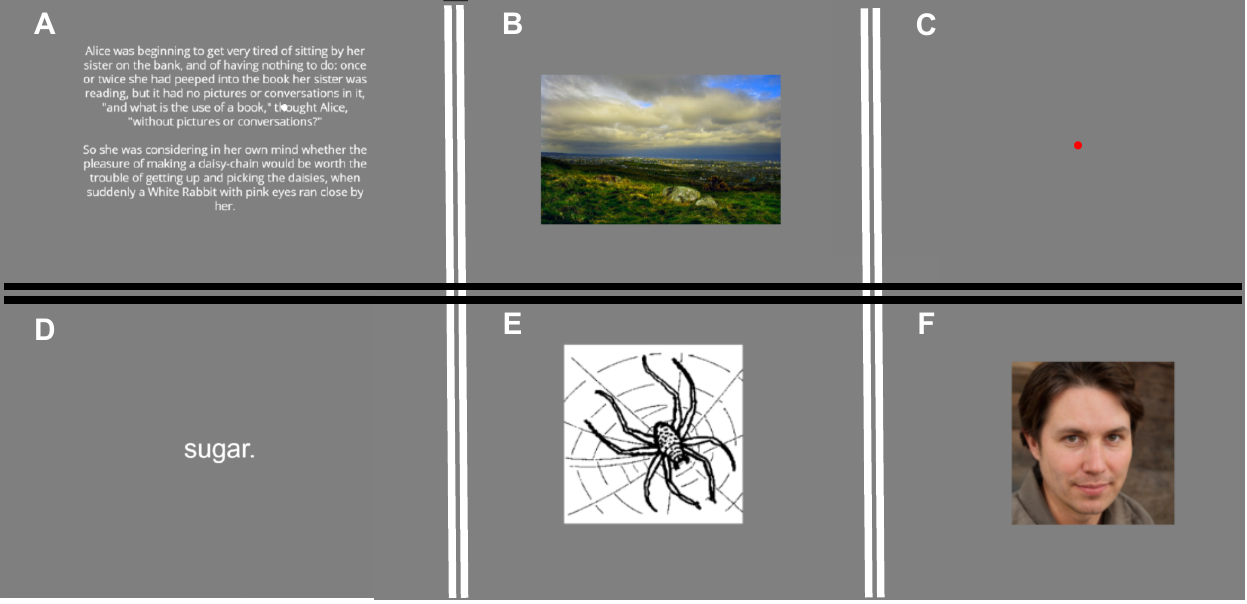}}}}
  \caption{ Left: Participant wearing the Emotiv EPOC X \cite{emotiv_epoc_x} neuroheadset while using our news website that required brainwave-based user authentication. Right: Interface screenshots of the authentication prototypes: A) Eye movement-Reading, B) Eye movement-Slideshow, C) Eye movement-Dot, D) Brainwaves-Reading, E) Brainwaves-Slideshow, and F) Brainwaves-Face.}
  \label{fig:mockup}
\end{figure*}

\textbf{Interfaces and Software Components.} For our study, we implemented interfaces for six authentication tasks, comprising three brainwave-based tasks and three eye movement-based tasks. To ensure consistency, we selected tasks with similar formats in both categories, e.g., based on looking at images or reading text. The final selection includes the best-performing tasks in the literature, according to their authentication accuracy, for which detailed information on their implementation is available. These tasks were either not subjected to usability testing in the original works, or were evaluated in a limited fashion (\Cref{sec:RW}). All tasks were developed using PsychoPy\footnote{\url{https://www.psychopy.org/}} and the interfaces are visually summarized in Figure~\ref{P:tasks}.

We implemented our \textbf{Brain Interface Prototypes} based on \textit{Arias-Cabarcos et al.'s} brainwave authentication experiment \cite{arias2021inexpensive,arias2023performance}, as follows:    

    \begin{itemize}
        \item The \textbf{Slideshow task} utilizes a technique that involves presenting an infrequent stimulus among a sequence of common ones, thereby generating a distinct and uniquely identifying brain response \cite{squires1975two}, often referred to as the oddball design. In our interface implementation, a specific photo is assigned as the target stimulus. Participants are presented with a sequence of images in which the target image shows up infrequently (20\% chance). Each photo is displayed for 200 milliseconds, followed by a random interval of 1 to less than 2 seconds before the next image appears. Furthermore, participants are instructed to count the targets, as this has proven useful in improving attention and increasing the amplitude of the brain signal of interest.

        \item The \textbf{Face task} is based on unique brain reactions that appear during face recognition, more specifically when observing an unfamiliar face after being primed with a series of familiar faces. Accordingly, our interface displays unfamiliar faces\footnote{ We used fake faces generated with Artificial Intelligence} amidst a stream of images featuring familiar faces (well established international celebrities), with an overall ratio of 1 unfamiliar face to every 3 familiar faces. 
        
        \item The\textbf{ Reading task} was designed to elicit uniquely identifying brain responses that appear in response to incongruent sentences. The interface shows sentences word by word, some of them ending in a semantically inconsistent manner \cite{kutas1980reading}. 
         \end{itemize}

For the \textbf{Eyetraking Interface Prototypes}, we base our implementations on the research works by \textit{Eberz et al.} \cite{eberz201928}, and \textit{Sluganovic et al.} \cite{sluganovic2018analysis}, as follows:
    \begin{itemize}
        \item The \textbf{Slideshow task }
        involves displaying a series of images to users in a slideshow format. Each image is displayed for a fixed duration of two seconds before being replaced by the next image in the sequence. This task is designed to assess eye-gaze authentication based on how the user's gaze moves and fixates on each image in the slideshow. 

        \item The \textbf{Dot task}
         displays a single red dot on a gray screen that changes position several times. Whenever the dot appears in the user's field of view, their reflexive ``saccades'' are triggered, causing a reorientation of their gaze towards the dot's new position. These reflexive eye movements have proven useful as a means of eye-gaze authentication. 

        \item The \textbf{Reading task} prompts the user to read a passage from the novel "Alice in Wonderland". The text is displayed in a central column on a grey background. Authentication is based on unique eye fixation features, which are captured while the users read the text.
        
\end{itemize}

\begin{table}[h!]
  \centering
  \begin{minipage}{.4\linewidth}
    \caption{Reported False Rejection Rates for \\Brainwave-based authentication tasks.}
    \label{tab:failBrain}
    \begin{tabular}{llc}
      \toprule
      \rowcolor{gray!30} \textbf{Mechanism} & \textbf{Task} & \textbf{FRR (\%)} \\
      \midrule
      \multirow{3}{*}{Brainwaves} & Slideshow (count) & 29.3 \cite{arias2023performance} \\
       & Face & 28.8 \cite{arias2023performance} \\
       & Reading & 38.5 \cite{arias2023performance} \\
      \bottomrule
    \end{tabular}
  \end{minipage}%
  \hspace{0.7cm}
  \begin{minipage}{.4\linewidth}
    \caption{Reported False Rejection Rates for \\Eyetracking-based authentication tasks.}
    \label{tab:failEye}
    \begin{tabular}{llc}
      \toprule
      \rowcolor{gray!30} \textbf{Mechanism} & \textbf{Task} & \textbf{FRR (\%)} \\
      \midrule
      \multirow{3}{*}{Eyetracking} & Slideshow & 12.37 \cite{eberz201928} \\
       & Dot & 19.84 \cite{sluganovic2018analysis} \\
       & Reading & 6.86 \cite{eberz201928}\\
      \bottomrule
    \end{tabular}
  \end{minipage}
\end{table}

In this study, we utilized interactive mock-ups for each task to simulate the real-world operations of the examined biometric techniques. It is critical to note that our decision to not deploy an actual biometric authentication system was motivated by two primary reasons: the current unavailability of reliable prototypes for brainwave and eye movement authentication and concerns over user privacy risks \cite{hanisch2021privacy, kablo2023privacy}. These limitations guided our choice to use simulated performance metrics, an approach that is consistent with established methodologies in prior biometric authentication user studies \cite{zimmermann2020password,khan2015usability, trewin2012biometric,rose2023overcoming}. We relied on performance data reported in the reference publications to configure the login failure rate, explicitly focusing on the False Rejection Rate (FRR), while setting the False Acceptance Rate (FAR) at 1\%. Although this FAR value is higher than current industry norms, it aligns with specific contextual constraints and anticipates future advancements in biometric technology. Tables \ref{tab:failBrain} and \ref{tab:failEye} provide a summary of the login failure rates or FRR for each authentication task, as reported in the literature.  We implement these failure probabilities in the developed prototypes.

\textbf{Hardware Components.} The experiment used two devices: 1) the Emotiv EPOC X\footnote{\url{https://www.emotiv.com/epoc-x/}}, a neuroheadset equipped with 14 electroencephalography electrodes (sensors) for brainwave data recording; and 2) the Tobii Pro Fusion\footnote{\url{https://www.tobii.com/products/eye-trackers/screen-based/tobii-pro-fusion}}, a compact screen-based eye tracker that can be plugged into a PC screen and captures gaze data at 128 frames per second. These devices were selected because they are higher-end consumer electronics, providing a good balance between accuracy, usability, and cost. Furthermore, this type of hardware has been used in the authentication research literature on which we base our prototype. Though we do not need the hardware for brain/eye data collection in our experiment, we use the devices to provide a realistic scenario, so the participants are convinced that the biometric mechanisms are fully implemented. Fig. \ref{F:emotiv} depicts a participant wearing the Emotiv headset while using our prototype \footnote{The participant provided consent fot he picture to be taken and published}.

%% file: 04_methodology.tex
\section{User Study Methodology}
\label{sec:Methodology}
We designed a lab study with an \textbf{interactive usage phase}, in which participants get to use the authentication prototypes, followed by a \textbf{post-usage survey}, to measure their perceptions. This section provides an overview of the experiment and  survey questionnaire design (\cref{sec:survey}), and describes the implementation, data analysis procedures, as well as the results of a pilot study (\cref{sec:data_analysis}). 

\subsection{Experiment Overview}

\begin{figure}[t!]
  \centering
  \includegraphics[width=\linewidth]{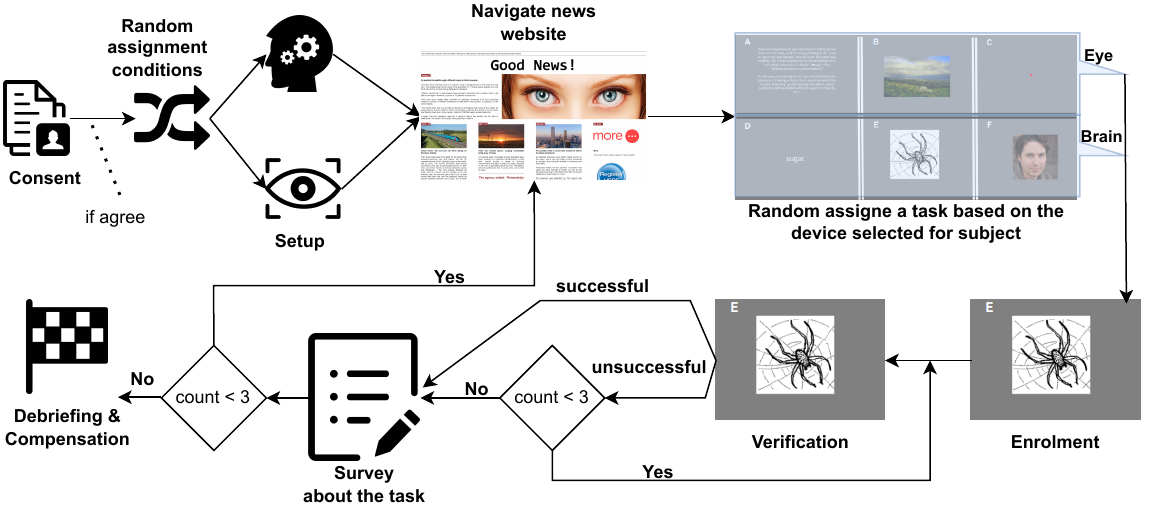}
  \caption{Detailed Experiment Process Flowchart. This figure illustrates the sequence of steps each participant follows in the study. Initially, participants sign a consent form, followed by a random assignment to one of the two authentication modalities: brainwave or eye movement. Subsequently, they undergo three separate enrollment and verification processes, each corresponding to one of the three authentication tasks within their assigned modality. After each task, participants are required to complete a short survey, making a total of three survey completions. The process concludes with a debriefing session and provision of compensation.}
  \label{P:eo}
\end{figure}

\label{sec:overview} 
Our study employs a between-subjects design for the two mechanism categories (brainwave and eye-tracking), and a within-subjects design for the three conditions under each mechanism. The experiment flow (Figure \ref{P:eo}) comprises these steps:\\

\textbf{Step 1 - Initialization.} The experiment begins by informing the participants about its goal: "\textit{testing new authentication systems}". Each participant is then provided with a comprehensive consent form approved by our university's Institutional Review Board (IRB) (Appendix \ref{SS:consent}). They are asked to read this form carefully, and if they agree with its terms, sign it to proceed with the experiment. Subsequently, participants are randomly assigned to either the brainwave or the eye-tracking authentication condition\footnote{Determined by a binary random process embedded in the main page of the experiment's local website.}. The device associated with the assigned condition is then set up and calibrated:

\begin{itemize}
\item \textbf{Emotiv EPOC X headset:} For this device, we involve participants in fitting the headset and adjusting the electrodes to achieve an optimal contact quality level. Although reaching 100\% connectivity can be challenging with participants with long or dense hair, we deem 80\% connectivity as an acceptable minimum after several minutes of attempting to improve it. This interactive process between researchers and participants adds a realistic touch to the experiment.

\item \textbf{Tobii Pro Fusion eye-tracker:} This device requires calibration, which participants conduct themselves. This involves adjusting their position to be within the camera's domain and focusing on a sequence of points on the screen. To enhance the realism of the experiment, we also provide participants with the opportunity to see a video of their pupils and gaze position on the screen.
\end{itemize}

The decision to assist with the setup of the EPOC X headset, specifically with electrode soaking process required to prepare the sensors, was taken to prevent unnecessary overhead for the participants and also due to the device's fragility. Furthermore, it is important to note that the landscape of technology in this area is evolving rapidly towards easier to fit headsets. Newer brainwave devices with dry electrodes are emerging that can be set up quickly and require minimal electrode adjustment\footnote{\url{https://www.bitbrain.com/neurotechnology-products/dry-eeg}}\footnote{\url{https://www.neurospec.com/Products/Details/1078/dsi-7}}\footnote{\url{https://www.emotiv.com/mn8-eeg-headset-with-contour-app/}}. Similarly, advancements in eye-tracking technology are leading to the development of calibration-free devices. Consequently, the setup process for these kinds of systems is becoming increasingly user-friendly, reducing the complexities previously associated with their use \footnote{\url{https://pupil-labs.com/products/invisible}}.

 \textbf{Step 2 - Interactive Usage Phase.} Once the device setup is successfully completed, participants are guided on how to navigate a news website displayed on a PC screen. Then, they are directed to a registration button to create an account and authenticate themselves, which allows them to access more comprehensive information about the news articles. Upon selecting registration, we initiate a task randomly based on the device assigned to the participant. This involves executing the enrolment and verification steps. The success rate for these verification attempts has been predetermined based on existing state-of-the-art literature, as detailed in Section \ref{sec:Prototypes}. If a login attempt fails, participants are allowed three attempts before we record it as a definitive login failure\footnote{This is a common practice implemented to restrict attacker success. After three failed attempts or one successful attempt, participants are directed to the post-usage survey (Appendix \ref{SS:survay}).}.

\textbf{Step 3 - Post-Usage Survey}. Participants fill out a digital survey designed to answer our research questions as follows:

\begin{itemize}
    \item 
\textbf{Usability (RQ1):} We assess the usability of the authentication systems using the established System Usability Scale (SUS) \cite{brooke1996sus}. It comprises 10 questions, each rated on a 5-point Likert scale. The scores for each question are transformed and aggregated to calculate an overall usability score ranging from 0 to 100. It is important to note that questions 4 and 10 measure a separate dimension related to Learnability, complementary to the overall usability concept (Appendix \ref{SS:A}).

\item \textbf{Perception and Usage (RQ2):} Participants are asked to provide their opinions and preferences regarding the presented authentication scheme. They have to rate their agreement with statements about the scheme's perceived security, ease of use, effort, and the balance between effort and benefits, following the approach of \textit{Zimmerman et al.'s} study \cite{zimmermann2020password}. Participants are also asked to indicate whether they would want to use the authentication mechanism if possible and for which types of applications and devices. In case they do not express interest in using the system, we ask for the reasons.Finally, they are requested to rank various authentication schemes based on their preferences (Appendix \ref{SS:B}). 

\item \textbf{Benefits, Problems, and Trade-offs (RQ3):} To get further insights, participants are asked open questions about the benefits and problems they see in using brainwave or eye movement authentication. Considering that time to authenticate and privacy are usually factors where users make trade-offs, we ask them to specify their acceptable authentication time, and to rate their level of agreement with a statement regarding their concerns about disclosing brainwave/eye movement data for authentication purposes \cite{zimmermann2020password} (Appendix \ref{SS:C}).

\item \textbf{Demographic Data and Background:} We collect basic information about the participants and their backgrounds. This section includes questions related to participants' prior experience with brain-computer interfaces, age group, gender, the highest level of education completed, and whether they have an educational/job background on IT. These questions provide valuable context for analyzing the survey results (Appendix \ref{SS:D}).
\end{itemize}

The interactive usage phase and post-usage survey, steps second and third, were repeated thrice for each participant to ensure comprehensive data collection across all three tasks associated with each device (Figure \ref{P:eo}). It should be noted that demographic and background questions (Appendix \ref{SS:D}) were only asked at the end of the first survey.

\textbf{Step 4 - Debriefing \& Compensation.} In order to ensure transparency, at the end of the experiment, participants are shown a page explaining that the prototype they interacted with was a simulation based on state of the art performance metrics. Participants are then asked if they had already realized that the experiment was not real and why. This question is useful to understand the validity of the experiment and to filter out participants who were aware of the artifact (Appendix \ref{SS:E}). Finally, participants were given compensation in cash as agreed in the consent form.

\label{sec:survey}

\subsection{Implementation and Analysis}
\label{sec:data_analysis}
\textbf{Recruitment \& Ethical Aspects.} 
Recruitment efforts were mainly focused on advertising through the official Instagram and Facebook accounts of Karlsruhe Institute of Technology (KIT), distributing flyers around the university, and utilizing email lists of students, targeting participants over 18 years old. 
Participation was voluntary and could be aborted at any time without negative consequences. 
The participant remuneration, fixed at 18 Euro ( for 75 minutes), slightly exceeds the minimum wage rate in Germany (12 Euro/hour)\footnote{\url{https://en.wikipedia.org/wiki/Minimum_wage_in_Germany}}. This higher compensation accounts for the necessity of participants' physical presence in the lab. Participants received this amount in cash. Participants' survey data was only analyzed for research purposes and handled in an anonymized way to ensure confidentiality. We used the SoSci Survey platform\footnote{\url{https://www.soscisurvey.de/}} for this purpose, as it is a flexible, GDPR-compliant survey platform. Since the tested behavioral biometric systems were simulated, we did not collect any biometric data from our participants. 
Before taking part in the study, all participants provided their consent. While we did hide the fact that the authentication prototypes were not real, we consider this slight deception as harmless: it does not put users at risk and it provides high benefits for research on ecologically valid scenarios. Our study was approved by the Institutional Review Board (IRB) of our university.

\textbf{Data Analysis.} SUS responses were analyzed using targeted hypothesis testing with $\alpha$ =.05
, selecting the appropriate test based on the data type and distribution. We used the non-parametric Friedman test for within-subjects testing and independent samples t-test for the between-subject case. All statistical analyses were performed using the R programming language. In order to gain a comprehensive understanding of individuals' experiences and contextualize quantitative results, we collected answers to open-ended responses. These responses were analyzed following an iterative, inductive coding approach~\cite{ miles1994qualitative}. One researcher developed the codebook with thematic codes after reviewing responses, while another independently coded the entire data. The inter-coder reliability, assessed using Cohen's Kappa \cite{cohen1960coefficient}, showed satisfactory agreement ($\kappa$ > 0.7). Discrepancies in codes were discussed and resolved. 
  
\textbf{Pilot Study.} We piloted the study with a small group of participants (N=3) to test the overall feasibility of design, assess the clarity of the instructions and questions, and evaluate the functionality of the software prototype.

One valuable insight we gained was that one of the participants easily recognized that the prototype was not real. He noticed poor electrode-to-brain connectivity during the setup phase, and despite their successful login, he realized that this could not work so the system was not authentic. Based on this feedback, we modified the protocol to ensure at least 80\% connectivity during the brainwave device setup. Additionally, for the eye-tracking mechanism, we focused on accurate calibration and allowed users to interact with their eye gaze on the screen for a few seconds, aiming to enhance their belief in the system's functionality.

As a result, during the debriefing phase of the main study, only three out of 35 participants (8.7\%) admitted that they already knew the system was not real. To understand their discernment, we inquired about the cues that led them to this realization. Participant S1 cited the \textit{“short process of determining a brain pattern”} as a clue, indicating skepticism towards the rapid authentication process. Participant S2's unfamiliarity with the system was evident as they remarked it was their \textit{“first time”} using such technology, suggesting a lack of prior exposure as a factor in their suspicion. Meanwhile, S3 simply perceived the setup as \textit{“just an experiment”} suggesting an inherent skepticism towards the authenticity of any setup within an experimental context. It is important to note that these responses do not point to a fundamental flaw in the experiment's design. Instead, they highlight individual experiences and perceptions, common in user studies involving both simulated and real systems. The high fidelity of our simulated system, therefore, remains credible, as these comments reflect isolated viewpoints rather than a collective assessment. Their surveys were subsequently excluded from our analysis to maintain the integrity of the data.

Additionally, the pilot study revealed some minor findings, such as ambiguity in some questions, typographical errors, the need for a country-specific layout for the keyboard, and a few technical issues. We addressed these concerns in the final version of the experiment.

%% file: 09_result.tex
\section{Results}
\label{sec:Results}
The study was conducted over a three-month period, from December 2022 to February 2023. A total of 35 participants took part in the study, of which 3 were filtered out because they realized about the simulation. From the final sample, 14 people used the brainwave-based authentication approach, and 18 used the eye-tracking-based authentication approach. 

\subsection{Participants Background}
The majority of participants (59.4\%) fell into the 18-24 age range, while 31.3\% were between 25-34, and 9.4\% were between 35-44. In terms of gender, 53.1\% identified as women, while 46.9\% identified as men. When considering educational achievements, 46.9\% of participants had a higher education level, with 28.1\% holding a Bachelor's degree, 15.6\% holding a Master's degree, and 6.3\%  with a Doctorate degree.

From the 33 participants, 68.8\%  had heard of eye-tracking devices and 43.8\% about brain–computer interfaces, but only 9.3\% in both cases reported that already have used these devices. The rest of the participants had no prior knowledge about the technologies.

Overall, the sample comprised a predominantly young and educated group of participants, with a slight gender imbalance favoring women. Limited familiarity with eye-tracking and brainwave devices was observed, as most participants reported only being aware of, but not used, these technologies. Full details of demographics per condition are given in Appendix Table \ref{tab:freq}.

\subsection{RQ1: Usability}
Usability results according to the System Usability Scale are presented in Table~\ref{tab:sus}. Brainwave authentication mechanisms got an average SUS of 79.6 ($\pm$ 14.3), slightly higher but very similar to the 78.6 ($\pm$ 13.3) obtained for the eye-tracking-based mechanisms. These scores are considered to be ``good'' (A$^-$), according to the qualitative grading from \textit{Bangor et al.}~\cite{bangor2009determining}  and  \textit{Sauro et al.}~\cite{sauro11are}. To gain deeper insight into the comparative usability of brainwave and eye-tracking mechanisms, we formulated three sub-questions:

\input{Table2}

\textbf{RQ1-1: Are eye-tracking-based authentication mechanisms more usable than brainwave-based mechanisms?} The distribution of SUS scores for the brainwave-based and eye-tracking-based mechanisms is visualized in Figure \ref{P:density}. To investigate potential differences, we conducted independent samples t-tests on the average SUS scores (Table \ref{tab:sus}), which indicated no significant difference exists.

\begin{figure}[h]
  \centering
  \includegraphics[width=\linewidth]{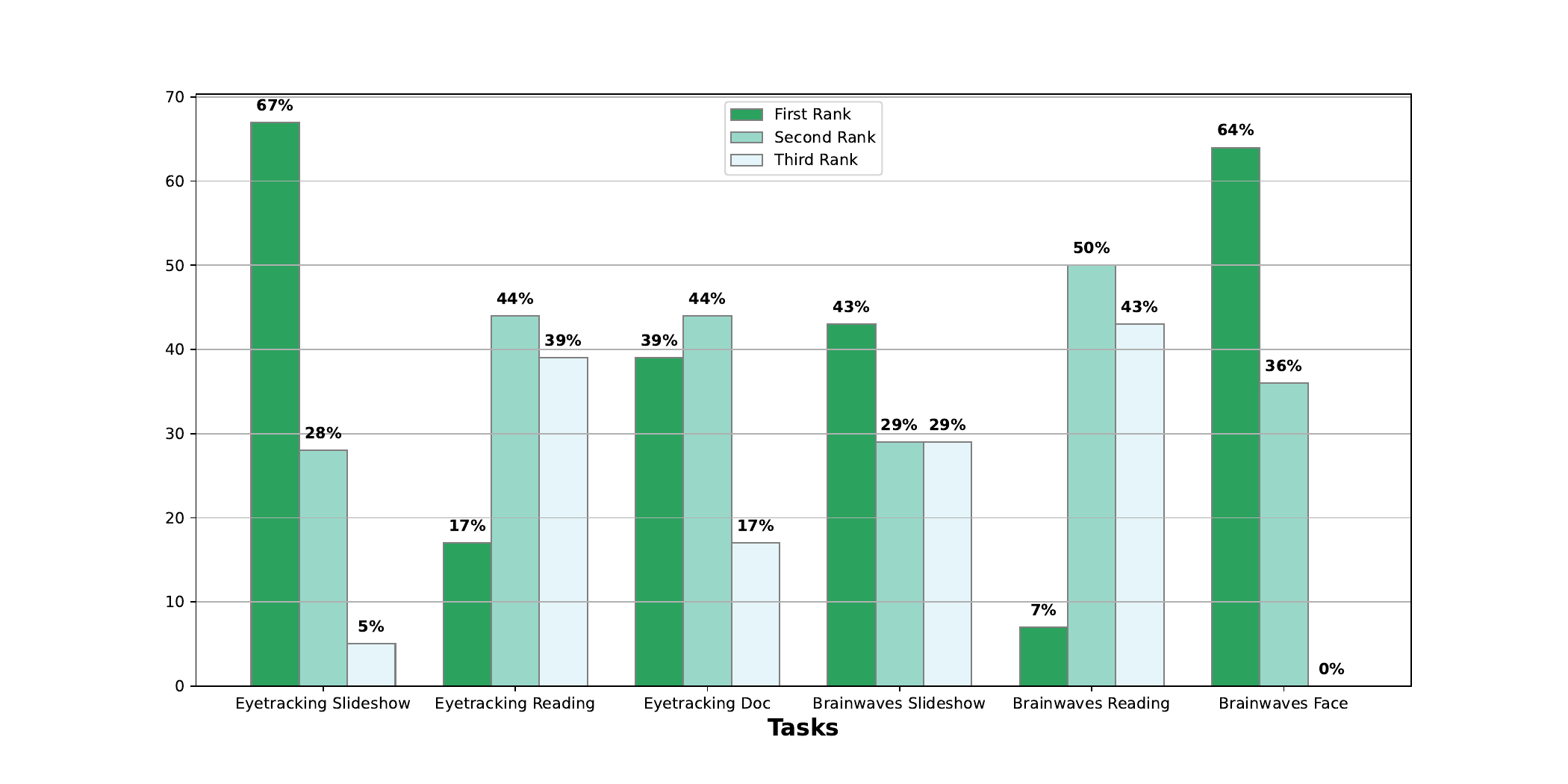}
  \caption{The plot displays the ranking percentages for each authentication task.}
  \label{P:Ranking}
\end{figure}

\textbf{RQ1-2: Which eye-tracking or brainwave-based authentication tasks are more usable?} The ranking of brainwave-based mechanisms from more to less usable was Face, with a SUS of 84.5 ($\pm$ 0.4), followed by Slideshow with 77.5 ($\pm$17.6), and Reading with 76.8 ($\pm$13.8). For the eye-tracking-based mechanisms, participants preferred the Slideshow (83.8 $\pm$11.5), over the Dot interface (78.8 $\pm$10.9), and the Reading task (73.2$\pm$15.4).

In assessing usability differences across interfaces, the Friedman tests indicated statistically significant disparities in usability scores, both in the brainwaves category (${{\chi}^2}(2)=9.2692$, p$=$.00971) and eyetracking category (${{\chi}^2}(2)=8.4$, p$=$0.015). The Friedman test, a non-parametric alternative to the ANOVA, is particularly suited for analyzing ordinal data across multiple groups. This prompted further investigation through Conover’s post-hoc test, employing a single-step p-value adjustment method to pinpoint specific task contrasts. Significant distinctions emerged, particularly between the Face and Reading tasks within the brainwave mechanism (p$=$.0065), and between the Slideshow and Reading tasks in the eye-tracking mechanism (p$=$.01).

To elucidate these findings, we transformed the System Usability Scale (SUS) scores into user-specific ranks. For instance, if a participant was assigned SUS scores of 81, 75, and 76, these were converted into ranks of 1, 3, and 2, respectively. This rank transformation, illustrated in Figure \ref{P:Ranking}, highlights the significant contrasts in user perceptions, especially underscoring the lower usability of Reading tasks. Only a minority of participants—17\% with eye-tracking and 7\% with brainwaves—rated Reading tasks higher than other evaluated tasks. Conversely, the Face task in the brainwave category and the Slideshow task in eye-tracking were deemed more user-friendly, receiving higher ranks from 64\% and 67\% of participants, respectively. These preferences align with prior research emphasizing the user inclination towards tasks with lower cognitive demand, such as those involving visual rather than textual stimuli ~\cite{arias2023performance,arias2021inexpensive}.

It is noteworthy that the SUS score for the Brainwaves-Face mechanism crosses the threshold of 84, which marks the difference from ``excellent'' usability to ``best imaginable'' (A$^+$), and the Eyetracking-Slideshow is close to it too~\cite{bangor2009determining}. These results reinforce the hypothesis that users generally favor interfaces that minimize cognitive and attention overhead.

\textbf{RQ1-3: Are eye-tracking and brainwave-based authentication mechanisms easy to learn?}
To gain further insights into usability, we also calculated the \textbf{learnability scores} for both categories of mechanisms: brainwaves and eye-tracking. The learnability scores revealed that they both were easy for participants to learn, with high mean scores of 88.7 and 91.0 for brainwaves and eye-tracking, respectively. Interestingly, eye-tracking showed a slightly higher mean score and lower standard deviation, suggesting it may be easier to learn and use overall. Among the authentication tasks, the Dot task had the highest mean learnability score (93.1) and lowest standard deviation (7.7), indicating it was the easiest task for participants to learn. Similarly, within the brainwaves mechanisms, the Face task had the highest mean score (91.1) and lowest standard deviation (15.8). However, nonparametric tests did not reveal statistically significant differences between the mechanisms or among the tasks.

\subsection{RQ2: Perception \& Usage}

\begin{figure}[t!]
  \centering
  \includegraphics[width=\linewidth]{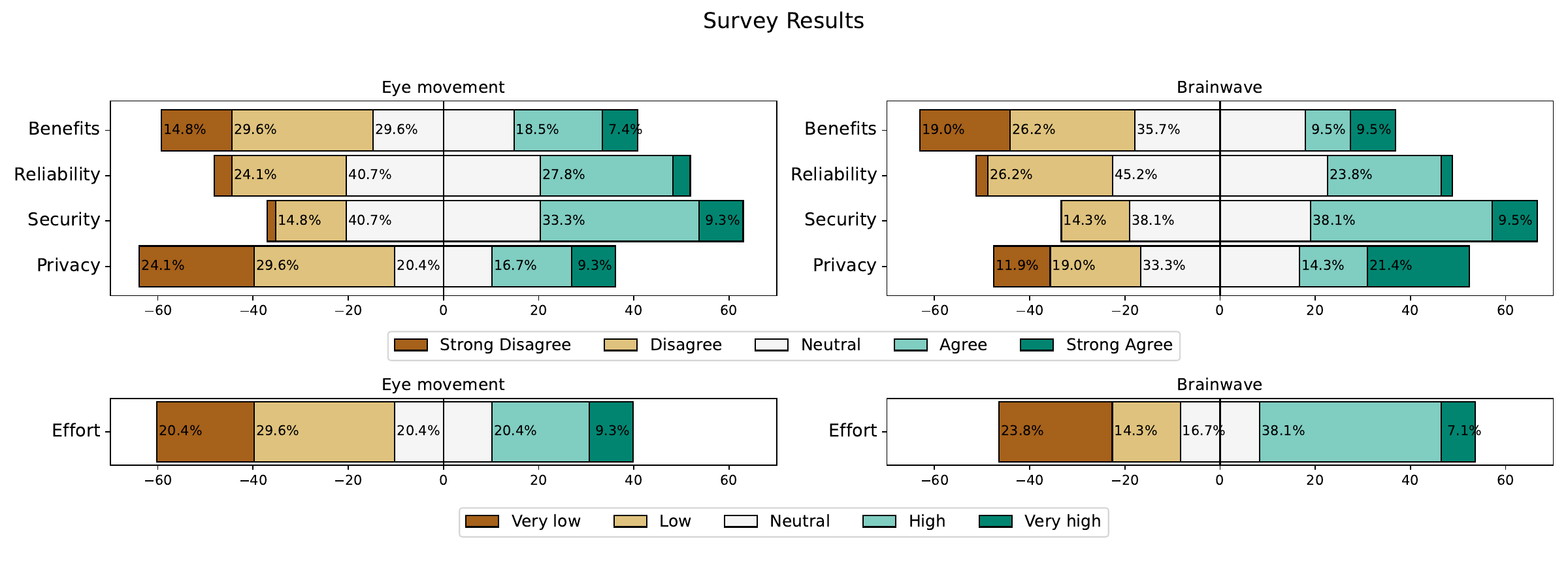}
  \caption{Subject perceptions on authentication scheme attributes for eye movement (left) and brainwave-based (right) mechanisms. Participants assessed: perceived Benefits (``\textit{In my opinion, the effort exceeds the gained benefits for this authentication scheme.}''), Reliability (``\textit{I think the use of this authentication scheme generally causes no problems}.''), Security (``\textit{I think this authentication scheme is very secure, that is, it protects me against attacks}''), Privacy concerns (``\textit{I have concerns to disclose eyegaze/brainwaves data for usage of an authentication scheme}.''), and Effort (``\textit{How do you rate the effort for using this authentication scheme?}'').}
  \label{P:pro}
\end{figure}

\textbf{Perceived Security, easy of use, effort, and reliability}
The results of the study indicate (Fig. \ref{P:pro}) that a slightly higher proportion of participants agreed or strongly agreed that the authentication scheme was very secure when using brainwave-based authentication mechanisms (47.6\%) compared to the eye-tracking mechanism (42.6\%). However, it is worth noting that a high proportion of participants chose a neutral answer in both mechanisms (38.1\%, 40.7\%). The relatively high percentage of neutral responses may suggest that participants had some uncertainty about the level of security offered by the authentication schemes, as highlighted in the qualitative analysis (\cref{sec:Qualitative}).

Regarding reliability, participants answered the question \textit{``}I think the use of this authentication scheme generally causes no problems\textit{''}. Interestingly, the results for both brain-based and eye movement-based mechanisms showed a similar trend, with a significant number of participants expressing neutrality (Fig. \ref{P:pro}). The findings indicate a lack of consensus on the absence of problems among the participants. 

In terms of user effort, participants were asked\textit{``}\textit{How do you rate the effort for using this authentication scheme?}\textit{''.} The results, as depicted in Figure \ref{P:pro}, indicate that 45.2\% of participants perceived the brainwave-based mechanisms as demanding high or very high effort, whereas only 29.7\% of participants reported similar perceptions regarding the eye movement mechanism. This suggests that brainwave-based mechanisms necessitate greater effort, potentially attributed to the additional overhead imposed by the headset used in this approach. Furthermore, participants expressed their opinions regarding the perceived balance between effort and benefits, as indicated in Figure \ref{P:pro}. 45.2\% and 44.8\% of the participants disagreed or strongly disagreed with the statement “\textit{In my opinion, the effort exceeds the gained benefits for this authentication scheme}'' for brainwaves and eye movement mechanisms, indicating a substantial level of disagreement rate. Notably, despite the higher effort required by the brainwave-based mechanisms, this category exhibited a similar disagreement rate compared to the eye-tracking mechanisms.

\begin{figure}[ht]
  \centering
  \includegraphics[width=\linewidth]{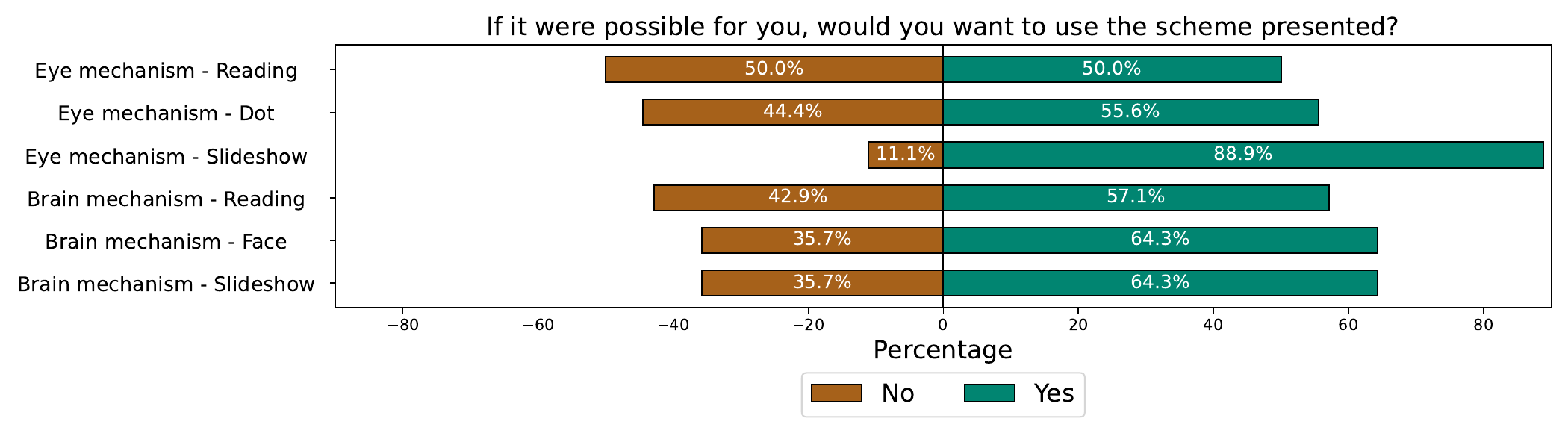}
  \caption{Participant Willingness to Use Different Authentication Schemes in Practice}
  \label{P:freq1}
\end{figure}

\textbf{Intended Usage.} To gauge the overall usage intention of participants, they were asked to indicate their willingness to use each authentication scheme by responding to the question: \textit{``If I had the possibility, I would use this authentication scheme''}. The results revealed that the brainwave mechanisms received an average response rate of approximately 62\%, while the eye movement mechanism had a response rate of around 65\%. These results indicate a notable interest in both these authentication schemes among the surveyed participants. Figure \ref{P:freq1} provides the frequency distribution of their responses, showcasing the alignment with the observed trend in System Usability Scale (SUS) scores. Particularly noteworthy is the overwhelmingly positive response of 88.9\% towards the Slideshow approach for eye-tracking. For participants who answered `No', we inquired about the reasons behind their decision via an open-ended question. The main reasons mentioned were security concerns (mentioned 10 times), authentication time (8), performance concerns (5), and the perceived burden of  using these authentication mechanisms (5). Additionally, participants in the Brainwave condition complained three times about the headset. Other reasons included a desire for more detailed information about the mechanisms (2) and a perception that the system is too complex.

\subsection{RQ3: Benefits, Problems \& Tradeoffs}
\label{sec:Qualitative}
\input{Table33}

To get richer insights into users' perspectives about the brainwave and eye-tracking authentication mechanisms, we utilized open questions aimed at identifying \textbf{benefits and problems}. In the following, we discuss the uncovered aspects for each category, by order of relevance (considering frequency of mention). The qualitative analysis is summarized in Tables \ref{tab:qp2} and \ref{tab:qp}.

\textbf{Benefits:} The most mentioned advantage was usability, which is backed up by the SUS scores in our quantitative analysis. Participants also value not having to deal with passwords and, interestingly, they mention improved security as a common positive aspect. The full list of topics that emerged contains:

\begin{itemize}
    \item \textbf{Usability.} Participants found both the eye-tracking and brain mechanisms to be easy to use. They expressed that these mechanisms were effortless and straightforward. Additionally, some participants specifically mentioned that these methods could be performed hands-free, further emphasizing the ease of interaction and convenience:
\vspace{0.2cm}
 \begin{displayquote}
    ``\textit{Very easy to use, and could be especially beneficial for example when trying to log into a service on a TV, it's a lot more comfortable than typing out your password with a TV remote}''- P26
     \end{displayquote}
\vspace{0.2cm}
   \begin{displayquote}   
``\textit{It can also prove handy when both the hands are already engaged and there is a need of authentication.}"- P10
     \end{displayquote}
\vspace{0.2cm}

    \item \textbf{Improve Security.} Several participants expressed a strong sense of security with both the eye-tracking and brain mechanisms. They viewed these authentication methods as highly unique and considered them to be potentially more secure than traditional methods such as fingerprints or regular passwords. Interestingly, the sense of security appeared to be even stronger in relation to brainwaves. They emphasized the difficulty of copying and stealing brainwave data: 

\vspace{0.2cm}
   \begin{displayquote}   
``\textit{It can be more secure than a fingerprint because when you are asleep no one can login to your device.''}- P18
    \end{displayquote}
\vspace{0.2cm}

    \item \textbf{Passwordless.} Some participants highlighted the advantage of the eye-tracking and brain-based mechanisms by noting that they eliminate the need to memorize passwords. They appreciated the convenience of not having to rely on remembering complex passwords and appreciated the alternative authentication approach offered by these methods. 
    \item \textbf{Authentication Time.} A few participants mentioned that they found this type of authentication to be fast. However, they did not provide specific details with regard to other authentication methods or elaborated on why they considered it fast.
    \item \textbf{Fun and Futuristic.} A few participants expressed that they found this authentication method to be fun and futuristic. They appreciated the unique and innovative nature of the approach, which added an element of excitement and novelty to the authentication experience.
\end{itemize}

\textbf{Problems:} the most salient problems that emerged in the analysis are:

\begin{itemize}
    \item \textbf{ Performance Concerns.} Several participants worried that the system may not be able to accurately track their eye movements, which could lead to problems such as difficulty logging in or inaccurate results as a potential obstacle to successful logins. Similarly, users are concerned that the brain mechanism may not be accurate enough to reliably identify them, leading to login failures. These performance-related concerns highlight the importance of reliability and consistency in system performance.
    \item \textbf{Mechanism Limitations. }During the study, participants expressed several concerns regarding the limitations of the proposed mechanisms due to their inherent nature. For the brain authentication mechanism, the requirement of wearing a headset emerged as a potential hurdle for widespread adoption. Participants pointed out that this additional hardware could be inconvenient and restrict their mobility. On the other hand, the eye-tracking mechanism also presented its own set of limitations. Participants highlighted the challenges faced by individuals with specific needs, such as dyslexic people, blind individuals, and those with visual impairments who rely on alternative methods for authentication. The mechanism's performance was also questioned in conditions such as dark environments and outdoor settings, where accuracy might be compromised. Furthermore, the presence of eyeglasses and sunglasses was found to obstruct or interfere with the accurate tracking of eye movements, adding another layer of complexity. Camera malfunctions were mentioned as a potential reliability issue for the eye-tracking system. 
    \item \textbf{Security Concerns.} Users expressed security concerns regarding both types of authentication mechanisms. Concerning brainwaves mechanisms, users highlighted a lack of trust in its security compared to traditional password-based authentication. They emphasized the need for robust security measures and questioned the system's ability to provide sufficient protection. In relation to the eye-tracking mechanism, users expressed even higher levels of security concern. They were worried about the ease of copying or recognizing their eye gaze data, raising doubts about its security. These security concerns underscore the importance of addressing user trust and ensuring the effectiveness and safety of both authentication methods.

    \item \textbf{Authentication Time.} Participants also raised concerns about the time required for authentication using the proposed schemes. Some participants felt that the system was not as fast as conventional methods such as typing passwords. Additionally, the possibility of repeated login attempts due to malfunctioning headsets was mentioned, indicating potential delays in the authentication process. These concerns emphasize the importance of efficient and swift authentication procedures to ensure user satisfaction.
     \item \textbf{Task Usability:} Participants reported encountering challenges related to task difficulty while utilizing the brainwave/eye-tracking scheme. Some participants expressed difficulties in concentrating on reading the text, indicating potential obstacles in effectively engaging with the system. Furthermore, a few participants found the flashing images uncomfortable, suggesting that the visual stimuli associated with the scheme may impede a seamless user experience. These findings underscore the significance of considering task demands and designing visual stimuli in order to optimize usability and minimize cognitive load for users interacting with the system. Additionally, concerns were raised regarding the mechanism's effectiveness for individuals who are illiterate or have difficulty reading. Moreover, A few participants expressed concerns regarding the potential exhaustion associated with frequent use of the system. The notion of the system becoming monotonous and boring over extended periods was mentioned, indicating the need to consider user engagement and system design elements that mitigate fatigue.
    
    \item \textbf{Privacy Concerns. }Privacy concerns emerged in the participants' responses, with a higher frequency for the brain mechanisms. Participants emphasized the importance of preventing the availability of their brain activity data to government entities and authentication providers, highlighting a desire to avoid any potential leaks or misuse of this sensitive information. The sensitivity of the collected brainwave data underscored the need for robust privacy measures and heightened awareness regarding the potential implications of unauthorized access or disclosure.

    \item \textbf{Cost of Equipment.} Some participants mentioned concerns about the cost associated with the required hardware. The perceived expense of the devices could pose a barrier to widespread adoption. Considering cost implications and exploring ways to make the system more affordable could enhance its accessibility and usability.
    \item \textbf{Transparency in Function.}  A few participants expressed uncertainty about how the system functions. They highlighted the need for clear explanations, particularly for non-technical individuals, to facilitate understanding and acceptance.
\end{itemize}

\textbf{Tradeoffs.} Previous studies identified time to authenticate and privacy as prominent factors influencing acceptance of novel behavioral biometrics~\cite{arias2021inexpensive, song2016eyeveri}. We investigate how these factors are perceived by potential users of eye-tracking and brain-based authentication mechanisms.

Regarding \textbf{time}, we asked participants `\textit{`What would be an acceptable/preferred authentication time?''}. Surprisingly, five out of 32 participants expressed a range of 1-5 minutes (one user mentioned 5 minutes, another mentioned 2 minutes, and three subjects indicated up to 1 minute) as acceptable authentication time. Two participants simply mentioned "a few seconds," which could not be quantitatively converted. For the remaining participants, the average preferred authentication time was 13.14$\pm$10.24 seconds. It is noteworthy that eight participants specifically emphasized a preference for a 5-second authentication time.

Regarding \textbf{privacy}, we asked participants whether they had concerns about disclosing their brainwave/eye-tracking data for the usage of an authentication scheme (see Fig. \ref{P:pro}). The results of the study indicate that a higher percentage of participants disagreed or strongly disagreed with privacy concerns about disclosing their eye-tracking data for authentication purposes compared to their brainwave data, which is an interesting finding. In fact, most of the participants, 54\%, who tried eye movement authentication are unconcerned versus 33\% of participants in the brainwave condition. In turn, 35\% of participants reported agreement or strong agreement to have privacy concerns in the brain case versus 26\% for eye movement. This difference in responses could be due to the fact that the brainwave device requires wearing on the head, while the eye-tracking device is non-invasive and attached to a desktop monitor. Another possible explanation is that participants believe that brainwave data may contain more private information drop when compared to eye-tracking data. Furthermore, the proportion of participants who selected ``neutral'' was highest for Brainwaves (33\%), compared to eye-tracking (20\%). This suggests that participants may have had more uncertainty or ambiguity regarding their privacy concerns when it comes to Brainwave data. It may be worth considering additional measures to elicit more nuanced responses in future studies, such as open-ended questions or follow-up interviews to better understand participants' perspectives.

%% file: Table2.tex
\begin{figure}[h]
    \centering
    \begin{minipage}{0.5\textwidth}
        \centering
        \captionof{table}{System Usability Scale (SUS) mean scores for Brainwave-based and Eyetracking-based authentication mechanisms. (*SD: Standard deviation)}
        \begin{tabular}{llccc}
            \toprule
            \textbf{Mechanism} &  \textbf{Task} & \textbf{Mean} & \textbf{SD*} & \textbf{Median} \\
            \midrule
            \multirow{3}{*}{Brainwaves} & Slideshow & 77.5  & 17.6 &  82.5\\
            & Face  & 84.5 & 10.4  & 87.5\\
            & Reading & 76.8&  13.8 &  78.8 \\
            \midrule
            \rowcolor{gray!30}  
            \textbf{Brainwaves} & \textbf{All} & \textbf{79.6} & \textbf{14.3} &  \textbf{82.5}\\
            \midrule
            \multirow{3}{*}{Eyetraking} & Slideshow &83.8 & 11.5 &  85  \\
            & Dot &78.8 & 10.9 &  78.8 \\
            & Reading &73.2 & 15.4  & 76.2 \\
            \midrule
            \rowcolor{gray!30} 
            \textbf{Eyetraking} & \textbf{All} &\textbf{78.6} & \textbf{13.3}  & \textbf{80} \\
            \bottomrule
        \end{tabular}
        \label{tab:sus}
    \end{minipage}%
    \hfill
    \begin{minipage}{0.45\textwidth}
        \centering
        \includegraphics[width=\linewidth]{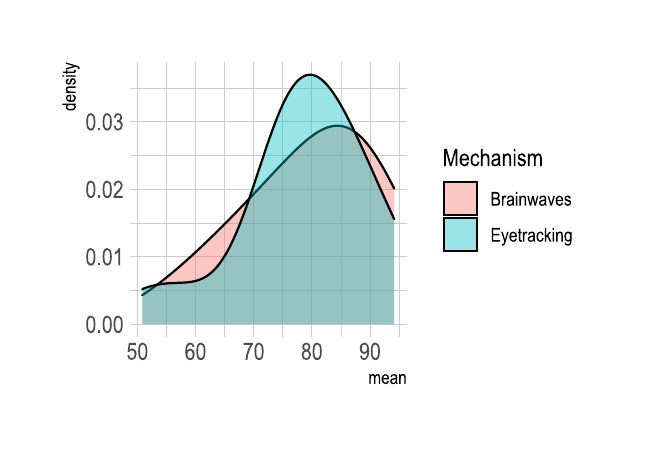}
        \captionof{figure}{The density plot shows the distribution of SUS scores of brainwave and eye-tracking mechanisms}
        \label{P:density}
    \end{minipage}
\end{figure}

%% file: Table33.tex
\begin{table*}
\small
\centering
  \caption{What benefits do you see about using a brainwaves activity/eyetracking authentication scheme?}
  \label{tab:qp2}
  \begin{tabular}{lp{11cm}c}
    \toprule
     \textbf{Codes} &  \textbf{Representative Quote}& \textbf{Ratio} \\

    \midrule
    
    \multirow{3}{*}{Usability} & \cellcolor{gray!20}\textit{Definition:} Ease of understanding and operation without extensive training. &  \\
     & \textbf{Eye:} ``It's easy to understand and you don't need to learn how to use it.'' & 34.59\%  \\
     &  \textbf{Brain:} ``I can see the benefit of building a more reliable and user-friendly authentication system.''& 38.05\%  \\\midrule

    \multirow{3}{*}{Improve Security} & \cellcolor{gray!20}\textit{Definition:} Enhancements in safeguarding data and access. &  \\
     & \textbf{Eye:} ``It seems to be more safe than a fingerprint or a face ID.''& 21.38\%  \\
     &  \textbf{Brain:}``The barrier to copying a password is very high. If it is only used locally for a password manager or similar, a lot of the attack surface vanishes.''& 33.63\%   \\\midrule

    \multirow{3}{*}{Passwordless} & \cellcolor{gray!20}\textit{Definition:} Elimination of traditional passwords, reducing the need to remember them. &  \\
     & \textbf{Eye:} ``The benefits for me are not having to worry about forgetting passwords.'' & 21.38\%  \\
     &  \textbf{Brain:}``Not having to remember a long secure password.''& 23.01\%  \\\midrule

    \multirow{3}{*}{Authentication Time} & \cellcolor{gray!20}\textit{Definition:} Concerns about the duration required for the system to authenticate a user. &  \\
     & \textbf{Eye:} ``It's quick to register and log in.'' & 8.81\%  \\
     &  \textbf{Brain:} ``It didn't take a lot of time to log in''.& 3.54\%   \\\midrule

    \multirow{3}{*}{Fun and Futuristic} & \cellcolor{gray!20}\textit{Definition:} Perception of the system as engaging and innovative. &  \\
     & \textbf{Eye:} ``It is a fun and entertaining way to enter a website (or else).'' & 11.95\%  \\
     &  \textbf{Brain:} ``It is fun, I feel like being in the future.''& 1.77\%   \\ \midrule

     \multirow{3}{*}{No benefit} & \cellcolor{gray!20}\textit{Definition:} users perceive no Problem. &  \\
     & \textbf{Eye:} ``i dont see any benefits. It gives me a headache.'' & 1.89\%  \\
     &  \textbf{Brain:} ``''& 0\%   \\
     
  \bottomrule
\end{tabular}
\end{table*}

\begin{table*}
\small
\centering
  \caption{What problems do you envision about using a brainwave activity/eyetracking authentication scheme?}
  \label{tab:qp}
  \begin{tabular}{lp{11cm}c}
    \toprule
     \textbf{Codes} &  \textbf{Representative Quote/Definition} & \textbf{Ratio} \\

    \midrule

    \multirow{3}{*}{Performance Concerns} & \cellcolor{gray!20}\textit{Definition:} Reliability and consistency of the authentication process. &  \\
    & \textbf{Eye:} ``Hardship in logging in due to inconsistent reading patterns.'' & 24.34\%   \\
    & \textbf{Brain:} ``There is no warranty you would be able to successfully login every time.''  & 20.61\%   \\\midrule

    \multirow{3}{*}{Mechanism Limitations} & \cellcolor{gray!20}\textit{Definition:} Issues related to practical use of the technology. &  \\
    & \textbf{Eye:} ``Eye-sight glasses and sunglasses could create hindrances...'' & 15.79\%   \\
    & \textbf{Brain:} ``I do not like to wear this headset...'' & 26.72\%   \\\midrule

    \multirow{3}{*}{Security Concerns} & \cellcolor{gray!20}\textit{Definition:} Potential risks regarding data safeguarding. &  \\
    & \textbf{Eye:} ``Why exactly it will be safer...'' & 18.42\%   \\
    & \textbf{Brain:} ``People might not trust brain activity based authentication...''  & 14.5\%    \\\midrule

    \multirow{3}{*}{Authentication Time} & \cellcolor{gray!20}\textit{Definition:}  Emphasizing quick and prompt user authentication. &  \\
    & \textbf{Eye:} ``Not fast enough compared to just typing a password'' & 14.47\%   \\
    & \textbf{Brain:} ``That brainwave headset wouldn't function properly...''  & 13.74\%   \\\midrule 

    \multirow{3}{*}{Task Usability} & \cellcolor{gray!20}\textit{Definition:} Challenges users face in interacting with authentication tasks. &  \\
    & \textbf{Eye:} ``Concentrating on reading a text was a little bit hard for me.'' & 14.47\%   \\
    & \textbf{Brain:} ``The images flashing is uncomfortable.'' & 7.63\%   \\\midrule

    \multirow{3}{*}{Privacy Concerns} & \cellcolor{gray!20}\textit{Definition:} Issues related to the handling and potential misuse of sensitive personal data. &  \\
    & \textbf{Eye:} ``Eye gaze data could be sensitive data'' & 4.61\%   \\
    & \textbf{Brain:} ``Information collected is sensitive and it may cause drastic problems if leaked.''  & 12.98\%   \\\midrule 

    \multirow{3}{*}{Cost of Equipment} & \cellcolor{gray!20}\textit{Definition:} Financial implications of implementing and using the technology. &  \\
    & \textbf{Eye:} ``For now expensive hardware.'' & 2.63\%  \\
    & \textbf{Brain:} ``I think the device would be too expensive.'' & 2.29\%   \\\midrule 

    \multirow{3}{*}{Transparency in Function} & \cellcolor{gray!20}\textit{Definition:} Understanding how the system operates and its mechanisms. &  \\
    & \textbf{Eye:} ``I dont know how it works.'' & 1.32\%  \\
    & \textbf{Brain:} ``Explaining the system to non-tech people could be more difficult.'' & 1.53\%  \\ \midrule

    \multirow{3}{*}{No Problem} & \cellcolor{gray!20}\textit{Definition:}  users perceive no Problem. &  \\
    & \textbf{Eye:} ``as long as the technical parts work I don't really see any problems at the moment'' & 3.95\%  \\
    & \textbf{Brain:} ``...'' & 0\%  \\

  \bottomrule
\end{tabular}
\end{table*}

%% file: 10_discussion.tex
\section{Discussion}
\label{sec:Discussion}
In this section, we provide recommendations for improving brain and eye-tracking-based authentication (\Cref{sec:recc}), give details on ecological validity (\Cref{sec:ev}), compare our results with the usability of other authentication mechanisms (\Cref{sec:comparative}), and discuss the limitations of our study \Cref{sec:limitations}. 

\subsection{Recommendations}
\label{sec:recc}
Our study findings support brainwave and eyetracking-based mechanisms as usable biometrics that users are interested to adopt. The SUS scores are located at the top of the scale signaling very good/excellent usability and surpassing the score of other mechanisms for which SUS have been reported in the literature (\Cref{sec:comparative}). While most of our participants report intention to use these mechanisms if available, our study highlights the need for improvements in three key areas to further enhance the usability and security of next-generation authentication mechanisms: privacy and transparency, task design, and verification time.

In particular, \textbf{privacy concerns} associated with the collection and usage of user data should be addressed. Our findings indicate that a non-negligible share of participants were uncomfortable with sharing their brainwave and eye-tracking data for authentication purposes. Therefore, designers and developers must ensure that user privacy is protected by implementing appropriate measures such as homomorphic encryption \cite{ccsmatin} or other techniques for biometric template protection, considering the overhead of implementing them in limited devices, such as neuroheadsets \cite{almenares2013overhead}. Additionally, it is advisable to process biometric samples on user devices, rather than on remote servers, especially in applications like XR authentication, and to clearly communicate the privacy benefits of such approaches to users. In terms of transparency, it is also important to educate users about how the device works and how the protection approach is implemented to alleviate their privacy concerns. Similarly, transparency should be provided in regard to the level of security offered, as many participants are uncertain or concerned about this. Adding this layer of transparency is an interesting future line of work to foster trust in novel biometrics.

\textbf{ Task design} can significantly impact the user experience of authentication mechanisms. In our study, we observed that the most popular tasks were the Face task from brainwaves and the Slideshow task from the eye-tracking mechanism based on SUS scores (Table ~\ref{tab:sus}). What is interesting about them is that both are easier compared to other tasks designed for these mechanisms. Specifically, tasks such as reading or counting require more attention from users and become less usable. Therefore, researchers and designers should aim to design visual and implicit authentication tasks, ideally embedded in users' usual activities, so they do not have to do something specific for authentication. Extra consideration is required to make the system accessible to people with different abilities, e.g., visual stimuli for brainwave-based authentication would not be suitable for blind people. Further research on other types of stimuli, such as sounds is desired to find alternatives that can be adapted to the user needs and usage context \cite{arias2019survey}. 

\textbf{Verification time} is also a critical aspect that requires attention. Our study indicates that some participants found the verification time to be too long, which could impact the overall usability of the authentication mechanism. Specifically, our results showed that most participants preferred an authentication time of 5 seconds or less, or at least the same as the time required to enter a password. Therefore, researchers and designers should aim to minimize verification time while maintaining an acceptable level of security.

Finally, our study found a significant percentage of neutral scores across different aspects of privacy concerns, perceived security, and expected problems with the authentication mechanisms. This suggests that a considerable number of users have neutral votes toward these factors and may not have strong preferences or opinions about them. However, it is important to note that even though some users may not express explicit concerns or preferences. To face this fact, further research can be conducted to investigate the underlying causes of these neutral scores.

As a final lesson learned with this study, we showed that building high-fidelity prototypes embedded in real-world scenarios works as a good proxy for early evaluation of behavioral authentication mechanisms. We recommend using this approach and incorporating feedback from
stakeholders early throughout the design process to ensure that
novel biometric technologies have the best chances for
achieving success.

\subsection{Ecological Validity}
\label{sec:ev}
Ecological validity refers to the extent to which research tasks and settings reflect real-world scenarios and can be generalized to real-life situations. In our study, we aimed to enhance ecological validity by incorporating several key elements into our experimental design. 

Firstly, we built a high-fidelity prototype that included a realistic new website browsing scenario, closely resembling typical online activities. This ensured that participants engaged with the authentication mechanisms within a context similar to their everyday experiences. Secondly, we implemented mechanism tasks that closely reflected those described in reference papers, ensuring that our study captured the essential elements of authentic authentication processes. Lastly, to further simulate real-world conditions, we incorporated authentication failure scenarios based on the failure rates reported in reference papers. This allowed us to mimic the challenges and frustrations that users may encounter during actual authentication processes.

It is worth noting that three out of 35 participants, or approximately 8.7\%, indicated that they were aware the system was simulated. We collected feedback on how they came to this realization: \textbf{S1} cited the `\textit{short process of determining a brain pattern}' as a clue. \textbf{S2} indicated unfamiliarity with the system, stating it was their `\textit{first time}' using it. \textbf{S3} simply noted that `\textit{it was just an experiment}.' We believe the high fidelity of our simulated system remains uncompromised based on these comments.

By considering these factors and incorporating them into our experimental design, we believe that our study possesses ecological validity, as it closely emulates real-world scenarios and provides meaningful insights into the usability and effectiveness of the authentication mechanisms.

\begin{figure}[t!]
  \centering
  \includegraphics[width=0.8\linewidth]{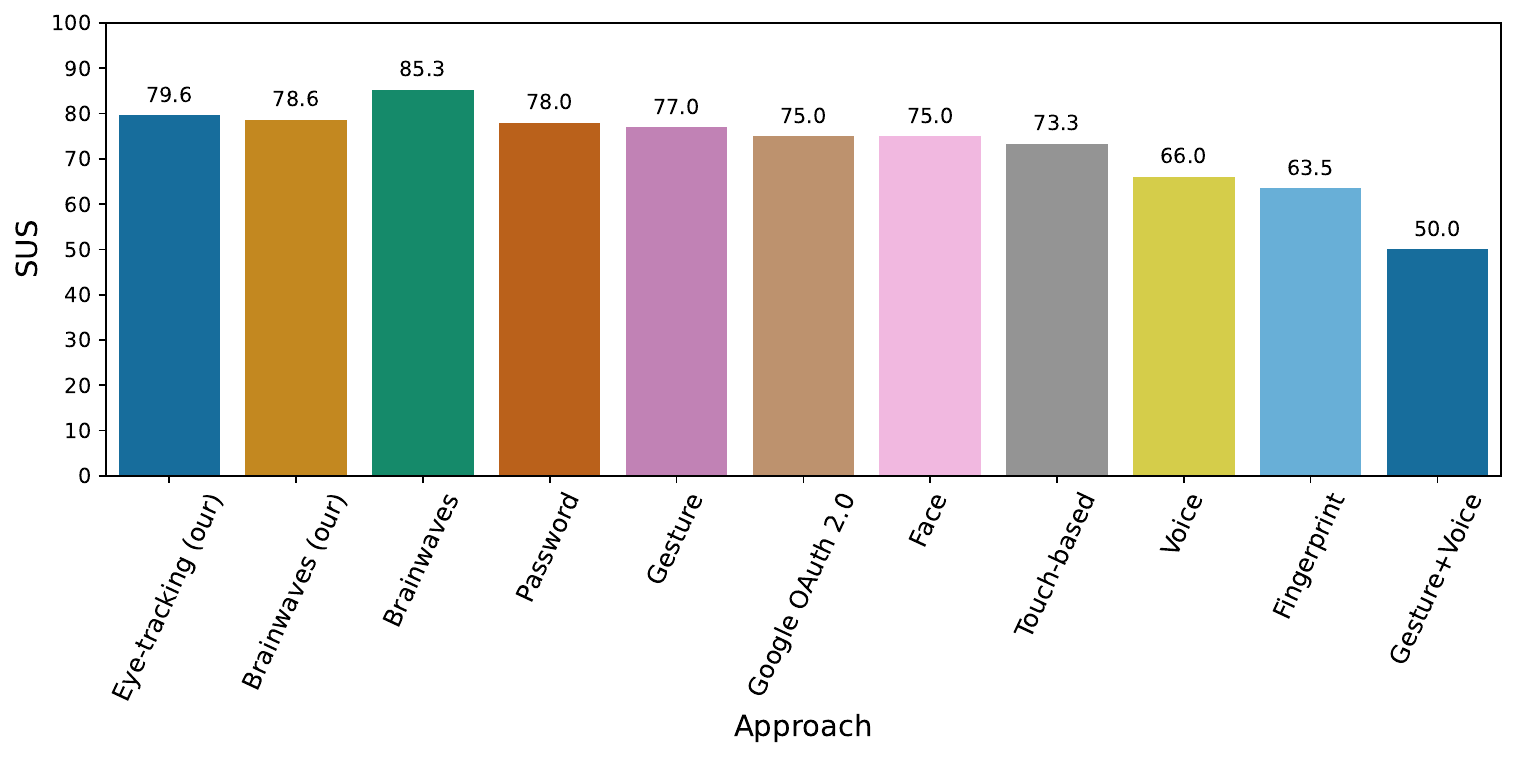}
  \caption{SUS scores for authentication with  Eye-tracking (our), Brainwaves(our), Brainwaves \cite{rose2023overcoming} 
  Password \cite{trewin2012biometric}, Gesture,
  \cite{trewin2012biometric}, Google OAuth 2.0 \cite{ruoti_authentication_2015},
  Face \cite{trewin2012biometric}, 
 Touch\cite{buriro2018dialerauth},
  Voice \cite{trewin2012biometric},
  Fingerprint \cite{oogami2020observation}, and Gesture+Voice
 \cite{trewin2012biometric}.}
  \label{P:susc}
\end{figure}

\subsection{Comparative Analysis}
\label{sec:comparative}
We compared the usability of our two mechanisms, eye-tracking and brainwaves, with eight different authentication mechanisms from other papers. We selected works that focused on biometrics and reported the System Usability Scale  (SUS) score. We also considered the evaluation of passwords and single sign-on (SSO) as a baseline.  The final set of authentication mechanisms available for comparison according to these criteria include: passwords, SSO (Google OAuth)\cite{ruoti_authentication_2015} \cite{trewin2012biometric}, as well as five biometric methods: Gesture \cite{trewin2012biometric}, face \cite{trewin2012biometric}, touch-based \cite{buriro2018dialerauth}, fingerprint \cite{trewin2012biometric}, and voice \cite{trewin2012biometric}. As it can be seen in Figure \ref{P:susc}, our eye-tracking and brainwave mechanisms achieved on average higher SUS scores than all of the other mechanisms. Comparatively, the traditional password-based authentication method obtained a SUS score of 78, demonstrating similar usability to our mechanisms. Furthermore, it is to note that most of our participants had no previous experience with the tested technologies and studies have shown that unfamiliar users rate new solutions 15-16\% lower in the SUS scale \cite{mclellan2012effect}.

When comparing the SUS scores with our work, it is important to consider that \textit{Trewin et al}.~\cite{trewin2012biometric} primarily focused on smartphone usability rather than the desktop scenario we examined in our study, and conducted the evaluation in 2012. Future work testing under the same conditions would be interesting to better assess comparative usability. The closest work, from\textit{ Ruoti et al}. \cite{ruoti_authentication_2015} compared seven non-biometric schemes in 2015, confirming that users prefer single sign-on vs the other secret and token-based solutions evaluated. Interestingly, their participants would like biometrics to be part of their ideal authentication system.

In our comparative analysis, we also consider the work of \textit{R\"ose et al}.~\cite{rose2023overcoming}, who reported an 85.28 System Usability Scale (SUS) score based on a brainwave slideshow task, though measured with a smaler sample (9 participants). Their findings closely align with ours, employing a similar survey process in their experiment. The slightly higher SUS score in their study could be attributed to the use of a more convenient and easy to wear brainwave recorder with only 4 dry sensors, the Muse 2\footnote{https://choosemuse.com/products/muse-2}. When assessing perceptions of security, 55\% of participants in\textit{ R\"ose et al.'s} study agreed or strongly agreed that their scheme was secure, compared to 47.6\% in our research. Regarding the ease of use, both studies observed a similar trend, with 55\% of their participants and 45.2\% of ours reporting neutral opinions. Additionally, when evaluating the benefit-effort trade-off, 55\% in\textit{ R\"ose et al.’s} study disagreed or strongly disagreed that the effort outweighed the benefits, closely paralleling our finding of 45.2\% disagreement. Finally, when asked about the willingness to adopt the scheme if available, 66.6\% of participants in their study responded positively, compared to an average of 61.9\% in ours.

\subsection{Limitations}
\label{sec:limitations}
The conducted study has some limitations that may impact the generalizability of the results. Specifically,  the evaluation was conducted with a relatively small sample size, and participants were recruited from a specific population (college students). Therefore, caution should be taken when extrapolating the results to other populations or contexts and further studies should be conducted to get insights from other populations. Additionally, the study only examined participants' initial attitudes towards disclosing their biometric data, and it is possible that attitudes could change over time with more exposure and education about the technology.

%% file: 11_Ack.tex
\section*{Acknowledgments}This work was funded by the Topic Engineering Secure Systems of the Helmholtz Association (HGF) and supported by KASTEL Security Research Labs, Karlsruhe, and Germany’s Excellence Strategy (EXC 2050/1 ‘CeTI’; ID 390696704).
This work was also partially supported by the Spanish Government under the research project “Enhancing Communication Protocols with Machine Learning while Protecting Sensitive Data (COMPROMISE)” PID2020-113795RB-C32, and the research project “QUantum-based ReSistant Architectures and Techniques (QURSA)” TED 2021-130369B-C32, both funded by MCIN/AEI/10.13039/501100011033.
\bibliographystyle{plain}
\bibliography{12_refrences}

\appendix
\section{Experiment Materials}
\subsection{Consent}
\label{SS:consent}
\subsubsection{Information}
\textbf{“Testing multimodal behavioral authentication”}\\
Thank you for participating in this study about new authentication mechanisms. We are researchers from the Computer Science
departments of ........... and the University of ..... In this study, we are investigating novel
mechanisms to verify users’ identity based on multi-sensor data, such as brainwaves and eye movements.
You can participate in this study if you are over 18 years old. The study consists of two main parts: i) a Usage phase, where you
navigate a website; and ii) a Survey phase, where you fill a questionnaire about the usage experience. The total duration of the
experiment is around 75 minutes, including breaks, and you will be compensated with 18 Euros in cash after completion.
The responsible researchers for this experiment are ............(email address)  and .............(email address) You can contact them if you have any questions about participation.
\subsubsection{Procedure and Participation}
If you decide to participate, you will first be introduced to the testing scenario, including the sensors used during the experiment.
Non-wearable sensors can be eyetracking devices or cameras attached to a PC, and you will be equipped with wearables if
necessary (e.g., with an electroencephalogram reader or a Virtual Reality (VR) headset). Then, you will conduct a website navigation
task. Afterwards, you will use a PC or tablet to answer a brief survey about your perceptions regarding the performed task, as well
as demographics and background questions.
The possible risks for the participants in this study are minimal, only those associated with computer tasks, such as mild fatigue or
slight discomfort from wearing the EEG or VR headset. The long-term benefits of this study for you are the potential outcomes of
this research that can aid in the development of better authentication technologies.
Participation in this study is voluntary and you are free to discontinue the study at any time. You will be compensated proportionally
to the time spent in the study.
\subsubsection{Data Collection and Processing}
Data collected from sensors will be used only during the testing task and removed right afterwards, we do not store any biometric
data. We only collect personal data in the survey; more specifically, the following demographics: age, gender, education level, and
technological background. The collection of socio-demographic data in the questionnaire is for the purpose of evaluating
heterogeneous groups. No attempt will be made based on the information you provide to draw conclusions about specific persons.
The evaluation results will be published in an anonymous form (tables and/or graphs), so that it is not possible to draw conclusions
about individuals.\\
($\square$ \small {I am over 18 years old, I have read and understood this consent form, and I agree to participate in the study})
\subsection{Post-Usage Questionnaire}
\label{SS:survay}
\subsubsection{Section A: Usability}
\label{SS:A}
\small
\textbf{System Usability Scale (SUS)}. Based on your experience, rate how much do you agree with the statements below?

\textcolor[rgb]{0.5,0.5,0.5}{(Answer choices:  \Square Strongly agree  \Square Agree  \Square Neither agree nor disagree  \Square Disagree  \Square Strongly disagree)}

\begin{itemize}
\item [SUS.01] \label{q58}I think that I would like to use this system frequently
\item [SUS.02] I found the system unnecessarily complex
\item [SUS.03] I thought the system was easy to use
\item [SUS.04] I think that I would need the support of a technical person to be able to use this system
\item [SUS.05] I found the various functions in this system were well integrated
\item [SUS.06] I thought there was too much inconsistency in this system
\item [SUS.07] I would imagine that most people would learn to use this system very quickly
\item [SUS.08] I found the system very cumbersome to use
\item [SUS.09] I felt very confident using the system
\item [SUS.10] \label{q67}I needed to learn a lot of things before I could get going with this system

\end{itemize}

\subsubsection{Section B: Perception and Usage}
\label{SS:B}
In the following, you will be asked questions regarding the presented authentication scheme. 
{\small
\begin{itemize}

    \item [B1] For each of the following statements, please rate how much you agree with it.(1 = Strongly disagree, 5 = Strongly agree)
    \begin{itemize}
        \item I think this authentication scheme is very secure, that is, it protects me against\textcolor[rgb]{0.5,0.5,0.5}{($\square 1 \:\square 2 \:\square 3 \:\square 4 \:\square 5$)}
        
        \item  I think the use of this authentication scheme generally causes no problems.\textcolor[rgb]{0.5,0.5,0.5}{($\square 1 \: \square 2 \: \square 3 \: \square 4 \: \square 5$)}

    \end{itemize}

    \item [B2] How do you rate the effort in using this authentication scheme?(1 = Very low, 5 = Very high)
      \textcolor[rgb]{0.5,0.5,0.5}{($\square 1 \:\square 2 \:\square 3 \:\square 4 \:\square 5$)}

    \item [B3] For the following statements, please rate how much you agree with it.(1 = Strongly disagree, 5 = Strongly agree)
    \begin{itemize}
 \item In my opinion, the effort exceeds the gained benefits for this authentication\textcolor[rgb]{0.5,0.5,0.5}{ ($\square 1 \:\square 2 \:\square 3 \:\square 4 \:\square 5$)}
    \end{itemize}    
    
    \item [B4] If it were possible for you, would you want to use the scheme presented? \textcolor[rgb]{0.5,0.5,0.5}{( $\square Yes \:\square No  $)}

    \item [B5] (if B4 == NO) Why wouldn’t you use this authentication scheme? 
    
    \item [B6] (if B4 == Yes) For which types of applications and services would you want to use the scheme?
      \textcolor[rgb]{0.5,0.5,0.5}{($\square Online \,Banking \:\square E-Mail \:\square Entertainment \,Websites \:\square Social \,Networks \:\square Apps \,provided \,by \,Health Provider \\ \:\square Apps \,at \,Workplace \:\square Government \,Administrative \,Services \\ \:\square Password \,Manager \:\square Other$)}
     
    \item [B7] (if B4 == Yes) In which devices would you like to use the presented scheme? \textcolor[rgb]{0.5,0.5,0.5}{( $\square  Smartphones \:\square Tablets \:\square TVs \:\square Computer \:\square Virtual\, Reality \,Headsets \:\square Gaming \,Consoles \:\square Digital \,Assistants (e.g., Amazon Echo)
    \:\square Other $)}

    \item [B8] (if round == 1) Which of these authentication mechanisms have you used?  \textcolor[rgb]{0.5,0.5,0.5} {($\square~Password \:\square~Fingerprint \:\square Face\, Detection \:\square~Signature \\ \:\square~Passwords\, with\, a \,Password\, Manager \:\square~Certificate \\ \:\square~Multifactor \:\square Key \,Strokes  \:\square~Graphical \,pattern  \\ \:\square~Yubikey \:\square~Other$)}
    
    \item [B9] (if round == 1) Rank these authentication mechanisms from most preferred to least preferred.(mechanisms already selected at B8) 
 
\end{itemize}}

\subsubsection{Section C: Benefits, Problems, and Trade-offs}
\label{SS:C}
This section contains questions designed to evaluate the benefits, problems, and trade-offs you associate with brainwave-based authentication. This data will provide us with valuable insight into your perception and will help to identify areas for improvement.
{\small
\begin{itemize}

    \item [C1] What benefits do you see about using this (brain or eyegaze) authentication scheme?
    \item [C2] What would be an acceptable/preferred authentication time?
    \item [C3] What problems do you envision with an authentication system that uses brain activity/eye gaze?
    \item [C4] Please indicate how strongly you agree with the following statement: (1 = Do not agree at all 5 = Fully agree)
    "I have concerns to disclose brainwave/eyegaze data for usage of an authentication scheme."
    \textcolor[rgb]{0.5,0.5,0.5} {($\square 1 \:\square 2 \:\square 3 \:\square 4 \:\square 5$)}

\end{itemize}}

\subsubsection{(if round == 1) Section D: Demographic Data and Background}
\label{SS:D}
This section includes a series of questions designed to collect information about the you as a person and your background. This
data will provide valuable context for the rest of the survey and help us to understand the results in the context of the overall
group of participants.
Reminder: All questions are optional. If you don't feel comfortable answering a question, you can simply skip it.
{\small
\begin{itemize}

    \item [D1] Which of the following describes your prior experience with "braincomputer interfaces"?\\Brain-computer interfaces allow direct control of a computer with your brain. The device you used today is such a device.\\
    \textcolor[rgb]{0.5,0.5,0.5} {($\square$ I \,had \,no \,knowledge \,of \,brain-computer \,interfaces  \,prior \,to \,participating \,in \,this \,study
    \:$\square$ I \,have \,heard \,about \,brain-computer \,interfaces  \,before \,participating \,in \,this \,study
     \:$\square$ I \,have \,already \,used \,a \,brain-computer \,interface  \,in \,the \,past
    \:$\square$ I \,own \,a \,brain-computer \,interface \,myself)}
    
    \item [D2] Which of the following age groups do you belong to?\textcolor[rgb]{0.5,0.5,0.5} {($\square 18 - 24 \,years \,old \:\square 25 - 34 \,years \,old\:\square35 - 44 \,years \,old\:\square45 - 54 \,years\, old\:\square55 - 64 \,years \,old\:\square65 - 74 \,years \,old\:\square75+ \,years \,old$)}
    
    \item [D3] What is your gender?\textcolor[rgb]{0.5,0.5,0.5} {($\square Female \:\square Male \:\square Non-Binary \:\square Other$)}
    
    \item [D4] What is the highest level of school you have completed or the highest\textcolor[rgb]{0.5,0.5,0.5} {($\square degree \,you \,have \,received? \:\square No \,Graduation \:\square Basic\, School\, Education \:\square Advanced \,School\, Education \:\square Higher \,Education \,Entrance  \,Qualification \:\square Apprenticeship \:\square Bachelor's \,Degree \\\:\square Master's \,Degree \:\square Doctorate$)}
    
    \item[D5] Which of the following best describes your educational background or
    job field? \textcolor[rgb]{0.5,0.5,0.5} { ($\square$ I \,have \,an \,education \,in, \,or \,work \,in, \,the \,field \,of \,computer \,science, \,engineering, \,or \,IT.
    $\square$ I \,do \,not \,have \,an \,education \,in, \,or \,work \,in, \,the \,field \,of \,computer \,science, \,engineering, \,or \,IT.)}
\end{itemize}}

\subsubsection{Section E: Debriefing}
\label{SS:E}
Thank you for your participation in this survey. We are investigating the acceptability and usability of emerging forms of digital authentication. These methods are not yet available, but there is research showing that they could work. In this experiment, we simulated the operation of a biometric authentication system for you to experience how it would look like, but there is no real system behind it, we didn’t actually register your biometric activity at all.

{\small
\begin{itemize}
\item [] Did you notice the authentication system was not real?\textcolor[rgb]{0.5,0.5,0.5} {($\square Yes \square No$)}
\end{itemize}}

\subsection{Additional Analyses}
\begin{figure}[t!]
  \centering
   \includegraphics[width=0.8\linewidth]{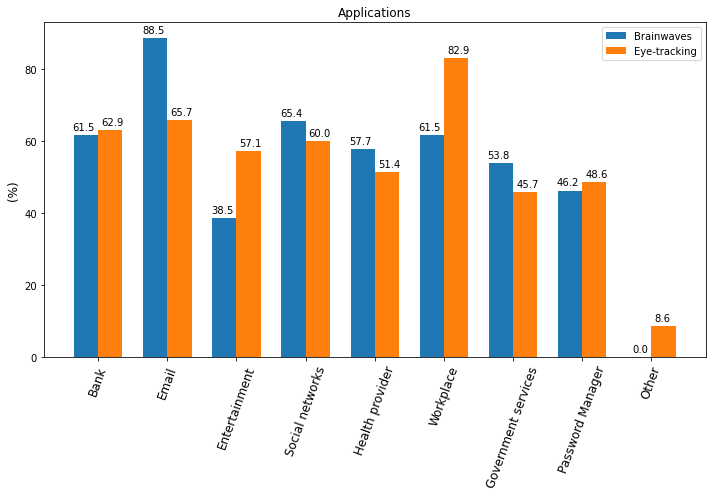}
  \caption{Answers to the question: ``For which applications would you use this authentication scheme''}
  \label{P:Contribution}
\end{figure}

\begin{figure}[t!]
  \centering
  \includegraphics[width=0.8\linewidth]{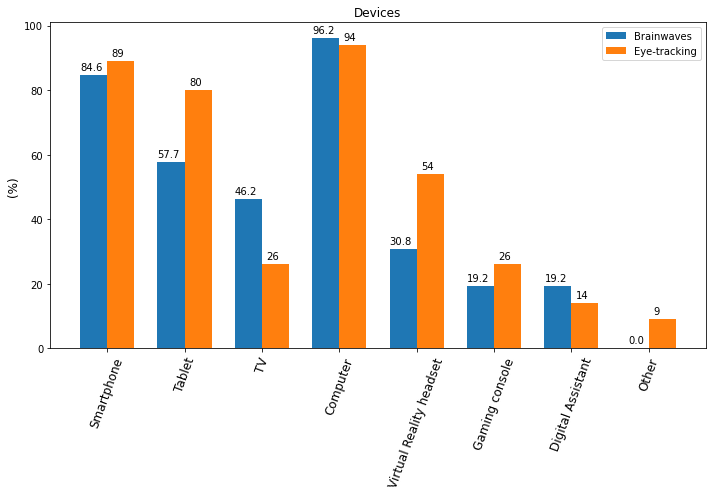}
  \caption{Answers to the question: ``On which devices would you like to use this authentication scheme? ''}
  \label{P:Contribution2}
\end{figure}

\textbf{Perception and Usage
.}
When participants expressed their interest in using each brainwave and eye-tracking authentication mechanism (Appendix \ref{SS:B}), we inquired about their intended applications for these authentication schemes. We asked them, \textit{`For which applications would you utilize this authentication scheme?'} Participants were instructed to select all applicable options. The corresponding results are presented in Figure \ref{P:Contribution}. The higher preference for eye-tracking compared to brainwave authentication may indicate concerns among participants regarding the utilization of brainwave mechanisms. Subsequently, we proceeded to inquire about the devices on which participants would prefer to employ this authentication scheme. The outcomes of this inquiry are depicted in Figure \ref{P:Contribution2}, revealing that participants exhibit a higher level of interest in using this authentication scheme on smartphones and computers.

\begin{figure}[t!]
  \centering
\includegraphics[width=0.8\linewidth]{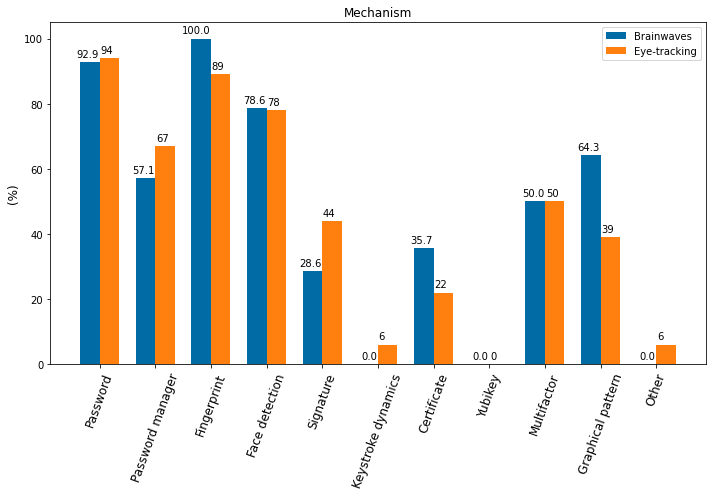}
  \caption{Answers to the question: ``Which of these authentication mechanisms have you used? (check all that apply)''}
  \label{P:Contribution3}
\end{figure}

\begin{figure}[t!]
  \centering
\includegraphics[width=0.7\linewidth]{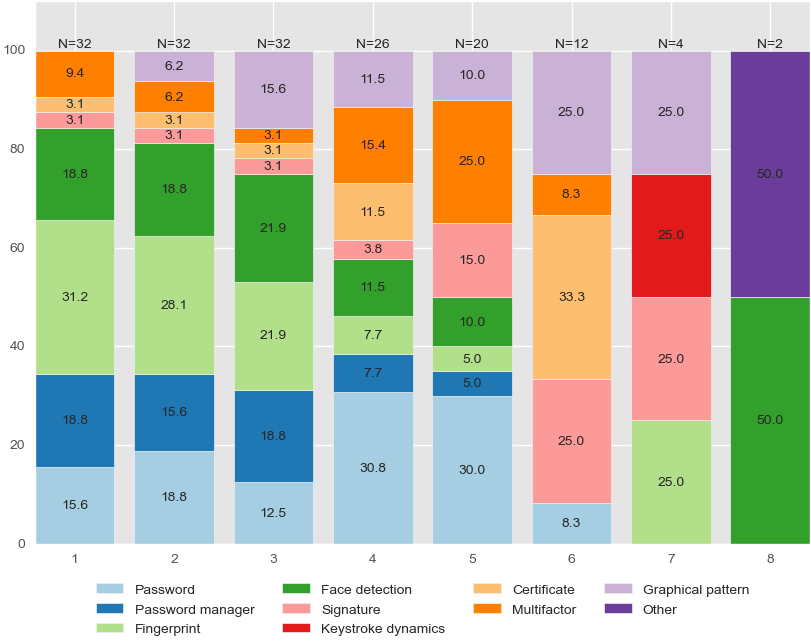}
  \caption{Answers to the question: ``Rank these authentication mechanisms from most preferred to least preferred.''}
  \label{P:rank}
\end{figure}

When participants were asked about the authentication mechanisms they had used (\ref{SS:B}), it was found that password, fingerprint, and face detection had a higher frequency of use (see Fig. \ref{P:Contribution3}). Furthermore, when participants were asked to rank these mechanisms, the results indicated that face detection and fingerprint were more popular compared to traditional passwords (see Fig. \ref{P:rank}).

 \textbf{Background Information:} Table \ref{tab:freq} provides comprehensive details regarding the background information of the participants.

\input{Table11}

%% file: Table11.tex
\begin{table*}
\small
\centering
  \caption{Study Participants' Background Information}
  \label{tab:freq}
    \begin{tabular}{@{}lccc@{}}\toprule
    ~                               & Brain Mechanisms(N=14) & Eye Mechanisms(N=18) & Overall(N=32) \\\midrule
\rowcolor{gray!30} \multicolumn{4}{c}{Preceding knowledge and experience with eye tracking devices}           \\\midrule
    No knowledge about eye tracking devices                    & 4 (28.6\%)            & 3 (16.7\%)          & 7 (21.9\%)    \\
    Heard about eye tracking devices                           & 8 (57.1\%)            & 14 (77.8\%)         & 22 (68.8\%)   \\
    Used some kind of eye tracking                             & 2 (14.3\%)            & 1 (5.6\%)           & 3 (9.4\%)     \\
    Own some kind of eye tracking devices                             & 0 (0\%)               & 0 (0\%)             & 0 (0\%)       \\\midrule
    \rowcolor{gray!30}\multicolumn{4}{c}{ Preceding knowledge and experience with Brain Computer Interface technology}         \\\midrule
    No knowledge about Brain Computer Interfaces                    & 6 (42.9\%)            & 9 (50.0\%)          & 15 (46.9\%)   \\
    Heard about Brain Computer Interfaces                           & 6 (42.9\%)            & 8 (44.4\%)          & 14 (43.8\%)   \\
    Used some kind of Brain Computer Interface                      & 2 (14.3\%)            & 1 (5.6\%)           & 3 (9.4\%)     \\
    Own some kind of Brain Computer Interface                       & 0 (0\%)               & 0 (0\%)             & 0 (0\%)       \\\midrule
    \rowcolor{gray!30}\multicolumn{4}{c}{Age}                                          \\\midrule

    35 - 44                         & 0 (0\%)               & 3 (16.7\%)          & 3 (9.4\%)    \\
    25 - 34                         & 3 (21.4\%)            & 7 (38.9\%)          & 10 (31.3\%)   \\
    18 - 24                         & 11 (78.6\%)           & 8 (44.4\%)          & 19 (59.4\%)   \\
   \rowcolor{gray!30} \multicolumn{4}{c}{Gender}          \\\midrule
    Woman                           & 8 (57.1\%)           & 9 (50.0\%)          & 17 (53.1\%)   \\
    Man                             & 6 (42.9\%)            & 9 (50.0\%)          & 15 (46.9\%)   \\
    \rowcolor{gray!30}\multicolumn{4}{c}{highest level of school you have completed or the highest degree}         \\\midrule
    Higher education entrance       & 8 (57.1\%)           & 7 (38.9\%)          & 15 (46.9\%)   \\
    Completed vocational training   & 1 (7.1\%)             & 0 (0\%)             & 1 (3.1\%)     \\
    Bachelor's Degree               & 3 (21.4\%)            & 6 (33.3\%)          & 9 (28.1\%)    \\
    Master's Degree                 & 1 (7.1\%)             & 4 (22.2\%)          & 5 (15.6\%)    \\
    Doctorate                       & 1 (7.1\%)             & 1 (5.6\%)          & 2 (6.3\%)     \\
    \midrule \rowcolor{gray!30}\multicolumn{4}{c}{Educational background or job field}            \\\midrule
    Education in, or work in, the field of computer science                     & 12 (85.7\%)           & 10 (55.6\%)         & 22 (68.8\%)   \\
    No education in, or work in, the field of computer science                  & 2 (14.3\%)            & 8 (44.4\%)          & 10 (31.3\%)   \\
    Prefer not to say               & 0 (0\%)               & 0 (0\%)             & 0 (0\%)       \\\midrule
   \rowcolor{gray!30}\multicolumn{4}{c}{English level}      \\\midrule
    B1 - Intermediate English       & 1 (7.1\%)             & 2 (11.1\%)          & 3 (9.4\%)     \\
    B2 - Upper Intermediate English & 7 (50\%)            & 4 (22.2\%)          & 11 (34.4\%)   \\
    C1 - Advanced English           & 2 (14.3\%)            & 10 (55.6\%)         & 12 (37.5\%)   \\
    C2 - Proficient                 & 4 (28.9\%)            & 2 (11.1\%)          & 6 (18.8\%)    \\\midrule
    \bottomrule
    \end{tabular}
\end{table*}